\newcommand\T{\rule{0pt}{2.6ex}}       
\newcommand\B{\rule[-1.2ex]{0pt}{0pt}} 
\documentclass{aa}  
\usepackage{graphicx}
\usepackage{natbib}
\usepackage{lscape}
\usepackage{subcaption}
\usepackage{multirow}

\bibpunct{(}{)}{;}{a}{}{,} 
\usepackage{txfonts}
\usepackage[colorlinks=true, citecolor=blue,linkcolor=red]{hyperref}
\begin{document}

   \title{Mid-infrared interferometry of 23 AGN tori: On the significance of polar-elongated emission}

   \author{N. L\'opez-Gonzaga\inst{1} \and L. Burtscher\inst{2} \and K. R. W. Tristram\inst{3} \and
  K. Meisenheimer\inst{4} \and M. Schartmann\inst{5}}

   \institute{Leiden Observatory, Leiden University, P.O. Box 9513, 2300 RA Leiden,
The Netherlands  \\ \email{nlopez@strw.leidenuniv.nl}
         \and 
         Max-Planck-Institut f\"{u}r extraterrestrische Physik, Postfach 1312,
         Gie\ss enbachstr., 85741 Garching, Germany
         \and
         European Southern Observatory, Alonso de Córdova 3107, Vitacura, Santiago, Chile
         \and
         Max-Planck-Institut f\"{u}r Astronomie, K\"{o}nigstuhl 17,
         69117 Heidelberg, Germany
         \and
         Centre for Astrophysics and Supercomputing, Swinburne University of Technology, P.O. Box 218, Hawthorn, Victoria 3122, Australia
             }

   \date{Received ;}

 
  \abstract
   {Detailed high resolution studies of AGN with mid-infrared interferometry have revealed parsec-sized dust emission elongated in the polar direction in four sources.}
   {Using a larger, coherently analyzed sample of AGN observed with mid-infrared interferometry, we aim to identify elongated mid-infrared emission in a statistical sample of sources.
   More specifically we wish to determine if there is indeed a preferred direction of the elongation and whether this direction is consistent with a torus-like structure or with a polar emission.   }
   {We investigate the significance of the detection of an elongated shape in the mid-infrared emission by fitting elongated Gaussian models to the interferometric data at 12 $\mu$m.
   We pay special attention to (1) the uncertainties caused by an inhomogeneous ($u, v$) coverage, (2) the typical errors in the measurements and (3) the spatial resolution achieved for each object. }
   {From our sample of 23 sources we are able to find elongated parsec-scale mid-infrared emission in five sources: three type 2s, one type 1i and one type 1. Elongated emission in four of these sources has been published before; NGC~5506 is a new detection.
   The observed axis ratios are typically around 2 and the position angle of the 12 $\mu$m emission for all the elongated sources seems to be always closer to the polar axis of the system than to the equatorial axis.
   Two other objects, NGC4507 and MCG-5-23-16 with a reasonably well mapped ($u,v$) coverage and good signal-to-noise ratios, appear to have a less elongated 12 $\mu$m emission. }
   {Our finding that sources showing elongated mid-infrared emission are preferentially extended in polar direction sets strong constraints on torus models or implies that both the torus and the NLR/outflow region have to be modeled together.
   Especially also models used for SED fitting will have to be revised to include emission from polar dust.}
   
   \maketitle

%

   \graphicspath{ {Figures/}   }

\section{Introduction.}

In Active Galactic Nuclei (AGN), a dusty toroidal structure is thought to be responsible for obscuring our line of sight onto the accreting Super Massive Black Hole (SMBH) for some inclination angles.
Obscured (``type 2'') AGNs, where only relatively narrow ($\lesssim$ 1000 km/s) forbidden emission lines are seen and unobscured (``type 1'') AGNs, which also show broad ($>$ 1000 km/s) permitted lines are then ``unified'' to the same class of objects simply by orientation \citep{1985ApJ...297..621A, 1993ARA&A..31..473A, 1995PASP..107..803U}.
This circum-nuclear dust, usually referred to as the ``dusty torus'', absorbs a fraction of the optical and UV light and re-emits the energy in the infrared regime.
The infrared bands are therefore a suitable window to analyze the structure of this dusty region.

Infrared observations at the diffraction limit of 10m class telescopes ($\sim$ 300 mas resolution) allow us to spatially isolate the emission of the circum-nuclear dust from the surrounding starburst emission, but do not usually resolve the structure of the ``torus'' \citep[e.g.,][]{2008A&A...488...83S,2008A&A...484..341R,2009A&A...502..457G,2010MNRAS.402..879R, 2009A&A...495..137H, 2009ApJ...702.1127R, 2011ApJ...731...92R, 2011A&A...536A..36A, 2012AJ....144...11M}.
Only in some 18 \% of objects, did \citet{{2014MNRAS.439.1648A}} detect extended emission on arcsecond scales.
In those cases, \citet{2014MNRAS.439.1648A} found that the extended emission usually coincides with the position angle of the Narrow Line Region, supporting the view that warm dust exists there \citep[see also][]{schweitzer2008}.

In recent years, infrared interferometry became an important tool to observe the dusty nuclear  regions of AGNs. 
With a resolution that can easily be more than 10 times better than that of a single-aperture telescope, this technique has allowed to resolve the cores of more than two dozen AGNs by now \citep[e.g.,][]{2004Natur.429...47J,  2009A&A...507L..57K, 2009A&A...493L..57K, 2009A&A...502...67T, 2013A&A...558A.149B}.
The analysis of the entire sample has shown that the mid-IR emission comes from a region about 4--20 $\times$ larger than the sublimation radius of dust, but also that the sample is very diverse in the sense that the size and structure is different in each object. 
To the extent that this allows statements about the average, no differences between type 1 and type 2 AGNs have been found.
For details we refer to \citet{2013A&A...558A.149B}, hereafter B13.

Previous interferometric studies of individual objects \citep[e.g.][]{2009MNRAS.394.1325R, 2012ApJ...755..149H, 2013ApJ...771...87H, 2014A&A...563A..82T, 2014A&A...565A..71L} on the other hand, are often able to resolve not only the basic structure, i.e. size and number of components, but also constrain the shape of the resolved emission. In the four objects that have been studied in detail, the dusty emission is always found along the polar direction rather than in an azimuthal configuration as one may naively expect from simple torus pictures \citep{1985ApJ...297..621A}.

The aim of this work is to coherently analyze the interferometric data of the MIDI AGN Large Programme (LP) sample (B13) to identify intrinsic elongations in the mid-infrared emission of the dusty region of the entire sample. 

The outline of this paper is the following: section 2 includes a brief explanation of the AGN sample and the interferometric data.
In Section 3 we explain our geometric model and the procedure followed to fit and identify  candidates with elongations. Our results are summarized in Section 4 and discussed in Section 5. We summarize our results in Section 6.

\section{The AGN sample, observations and data processing}

Here we use the sample first presented in B13. 
It consists of the MIDI AGN Large Programme and all other extragalactic mid-IR observations publicly available until the time of publishing. These are 23 Active Galactic Nuclei, all observed with the MID-infrared interferometric Instrument \citep[MIDI, ][]{2003Ap&SS.286...73L} at the European Southern Observatory's (ESO's) Very Large Telescope Interferometer (VLTI) on Cerro Paranal, Chile.
MIDI is a classical Michelson interferometer combining the beams of two telescopes at a time in the atmospheric $N$ band, i.e. the wavelength region 8 -- 13 $\mu$m.
Due to the high background in the thermal infrared (exacerbated by $\sim$ 20 reflections along the VLTI optical train), severe flux limits apply. 
Our sample is therefore essentially flux-selected among the brightest and most well-known AGNs. 
It includes twelve type 1, nine type 2 Seyfert galaxies, one radio galaxy and a quasar, with a median luminosity distance of 53 Mpc. For this work, we additionally collect the published data for the Circinus~galaxy \citep{2014A&A...563A..82T}, NGC~1068 \citep{2014A&A...565A..71L} and NGC~3783 \citep{2013ApJ...771...87H}, three galaxies whose interferometric data showed clear signs of elongated emission.
For details about the observation sequence and data reduction, we refer to \citet{2012SPIE.8445E..1GB}.

All sources were observed with pairs of 8 m Unit Telescopes (UTs) in at least 3 different baseline configurations.
Additionally, for the two  brightest and closest AGNs, Circinus and NGC1068, observations were also carried out using configurations with the 1.8 m Auxiliary Telescopes (ATs) to cover shorter baselines.
These two sources are almost completely resolved out in the interferometric data, while in the other galaxies less than 60\% of the flux is resolved out.
The reason is that the relative resolution, with respect to the dust sublimation radius, is much higher in these two galaxies than for the rest of the sources in the sample.
In order to achieve a sample with homogeneous spatial resolution (in multiples of the innermost radius of dust), we therefore only use the interferometric data obtained from projected baselines $< 40 $m for the two bright sources.

MIDI also provides limited spectral information in the $N$ band.
This information can be used to study several physical properties of the torus, such as temperature profiles or the nature of the silicate feature.
As the purpose of our study is to investigate the general geometrical shape of the continuum  mid-infrared emission, we restrict ourselves to the 12 $\mu$m (rest frame) fluxes, as it was done in B13. 
At this wavelength, the sources are best resolved and observations are least affected by instrumental and calibration errors, such as correlation losses, the silicate feature or atmospheric absorption. 

The lack of true phase information from the instrument and the sparse ($u,v$) coverage due to the time-consuming observations, do not allow us to apply image reconstruction techniques.
Instead, we can only forward model simple geometrical brightness distributions and compare them with the observed visibilities.
These models need to be kept as simple as possible to avoid degeneracies. While the true brightness distribution of the objects might be quite complex, it has been shown that simple analytical surface brightness distributions, such as Gaussian or uniform disks, can be used to provide a first order approximation of the shape and size and serve as building blocks for more complex geometries \citep[see e.g.,][]{2004Natur.429...47J, 2006A&A...450..483P, 2009ApJ...705L..53B, 2013A&A...558A.149B}.

\subsection{Averaging adjacent ($u,v$) points}

Before modeling the interferometric data, we replaced measurements of adjacent ($u,v$) points with a weighted average.
According to \citet{2014A&A...565A..71L}, if the size of the emission is smaller or similar to the resolution of the single-aperture telescope, measurements taken with adjacent ($u,v$) points should not vary too much if the distance $\Delta u$ between them is $\Delta u \lesssim$ D, where D is the diameter of the telescope.
Since all of our objects are essentially unresolved with the single-aperture telescope, we average ($u,v$) points with $\Delta u \lesssim 8$ m, except for NGC1068 and Circinus where we use a $\Delta u \lesssim 1$ m as they were observed with the ATs and they show significant extension at arcsecond resolutions.

Points within a distance $\Delta u$ from each other are average together using their corresponding uncertainty as a weight.
Because the actual changes over the ($u,v$) plane in terms of correlated flux are small compared to our measurement error, we can assume that the differences in adjacent points is mostly dominated by noise.
The averaging process not only serves to get more precise visibilities, but it also serves as a way for re-binning the ($u,v$) plane in order to reduce the non-uniform sampling.

\section{Finding elongations.}

\subsection{Elongated model}

\begin{figure*}
\centering
\sidecaption
	\includegraphics[trim=0cm 8cm 0cm 0cm, width=0.95\hsize]{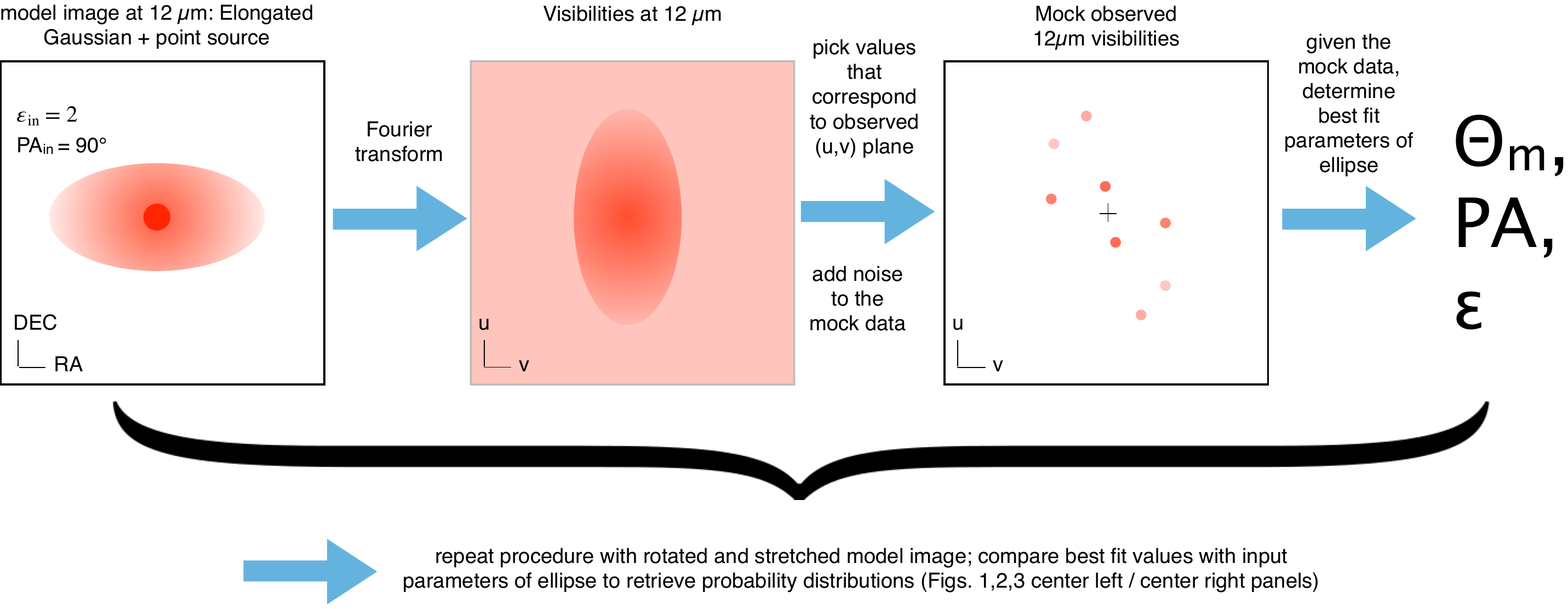}
	\caption{\label{fig:flowchart} Outline of the test setup to determine the reliability of our model fit for each source. First we generate a simple model image (left panel) consisting of a point source and an elongated Gaussian of arbitrary axis ratio and position angle. Since the intensity of the point source is largely model-independent and well constrained by the observations, its visibility can be set to the actual value of the respective source. Next, this ``image'' is Fourier transformed (center left). From these model visibilities we take the values that correspond to the actually observed $(u,v)$ coverage for the specific source (center right panel). After adding appropriate noise, we fit the mock data set with our three-parameter fit (minor axis FWHM $\theta_m$, position angle PA and axis ratio $\epsilon$) and compare the result with the known input parameters. This procedure is repeated for different axis ratios and position angles and for the specific ($u,v$) coverage of each source.}
\end{figure*}

Previous results from B13 using symmetric distributions, allowed to determine an average size of the infrared emission for the AGN tori of our sample.
But from the residual plots in Figures 5-27  of B13 it was, however, already clear that for some objects the data contain more information about the shape than just an average size.

The 12 $\mu$m emission of the large program sources observed with MIDI can be described with one or two components: 1) an unresolved emission and sometimes 2) a (over)resolved  component. 
For every object the level of unresolved emission (the minimum visibility of point source fraction) is well determined even without any model fitting in most of the sources (see B13), this emission is best described by a point source distribution. If there is evidence for a second component we can describe it by using elongated Gaussian component as it is done for this work (the circular Gaussian component used in B13 is a particular case of the elongated model).

The correlated fluxes are defined by the following equation 
\begin{equation}
F_{corr} (u,v) = (F_{tot}-F_{pt}) V_{gs}(u,v)+ F_{pt}
\end{equation}
where $F_{corr}$ is the correlated flux, $F_{tot}$ is the total flux measured by the single-aperture telescope, $F_{pt}$ is the flux of the point source and $V_{gs}$ is the visibility of the elongated Gaussian distribution.
The elongated Gaussian distribution is defined by three parameters: the Full Width at Half Maximum (FWHM) of the minor axis ($\theta_m$), an axis ratio $\varepsilon= \Theta_M/\theta_m$ defining the ratio between the FWHM of the major axis ($\Theta_M$) and the FWHM of the minor axis ($\theta_m$), and the position angle ($PA$) of the major axis measured in degrees from east of north.
The analytical description of the correlated fluxes is given by the following formula,
\begin{equation}
 V_{gs}(u,v) = \exp \left[ -\frac{\left(\pi\, \theta_m \sqrt{u'^2+ \varepsilon^2 v'^2}\,\right)^2}{4 \ln 2}\right]
\end{equation}
where $u'=u \cos(PA)-v \sin(PA)$ , $v'=u \sin(PA)+v \cos(PA)$.
For most of our objects, the total 12 $\mu$m flux has a relatively low uncertainty so we ignore this error and for the rest of the paper we will consider the total single-aperture flux as a fixed quantity and treat the correlated fluxes as visibilities.


\begin{figure*}
	\includegraphics[width=0.9\hsize]{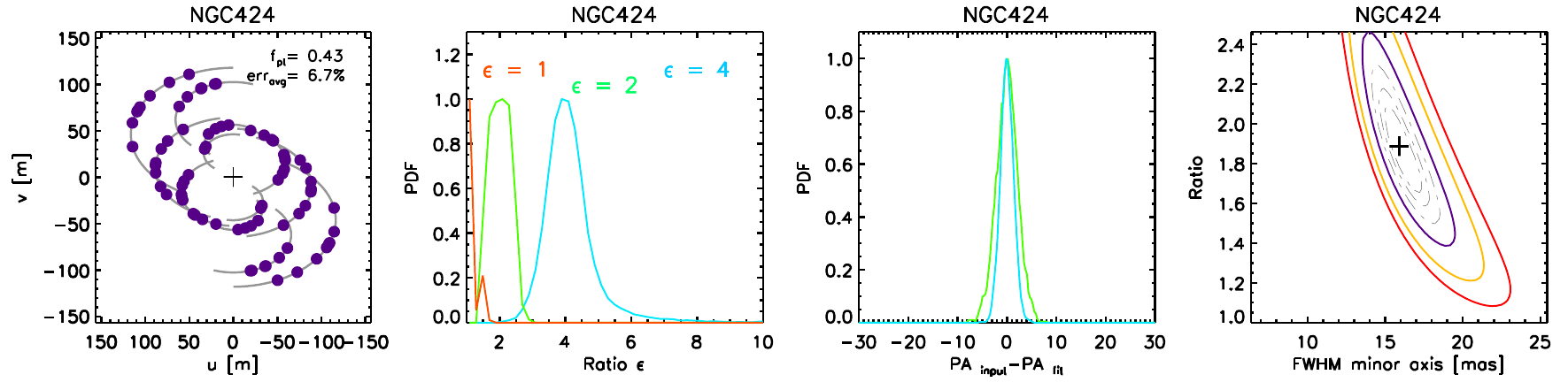}\\
	\includegraphics[width=0.9\hsize]{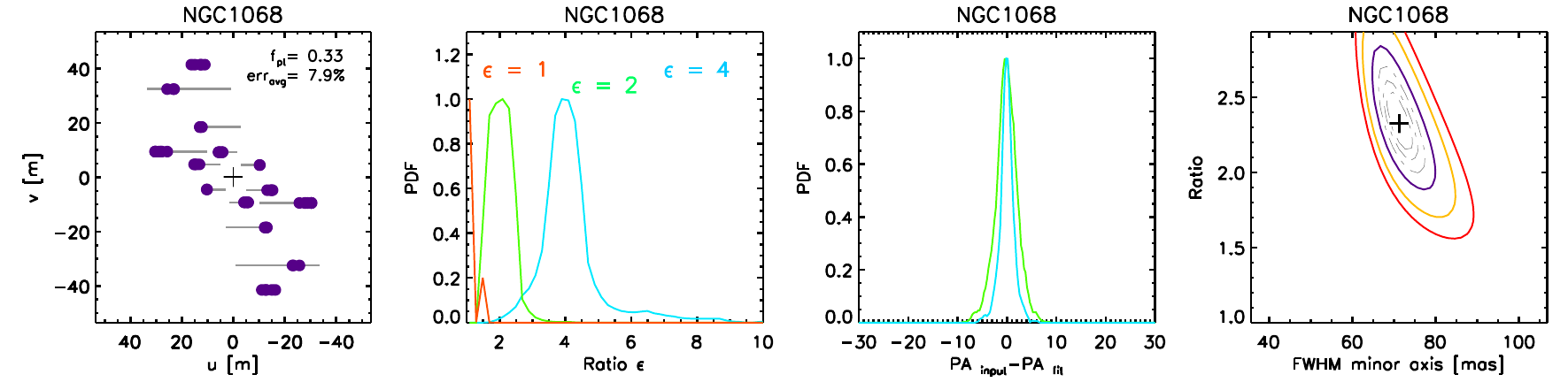}\\
	\includegraphics[width=0.9\hsize]{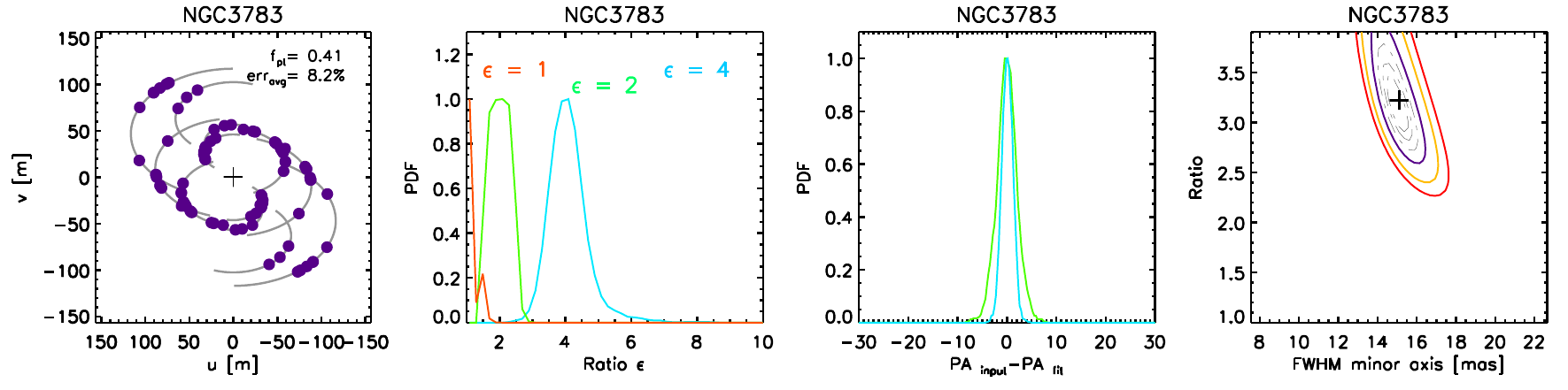}\\
	\includegraphics[width=0.9\hsize]{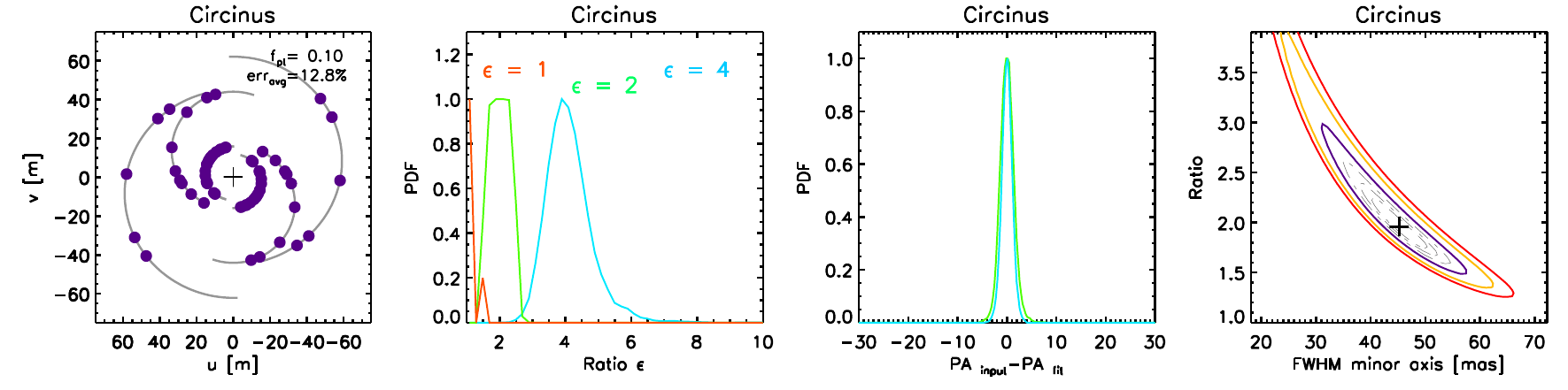}\\
	\includegraphics[width=0.9\hsize]{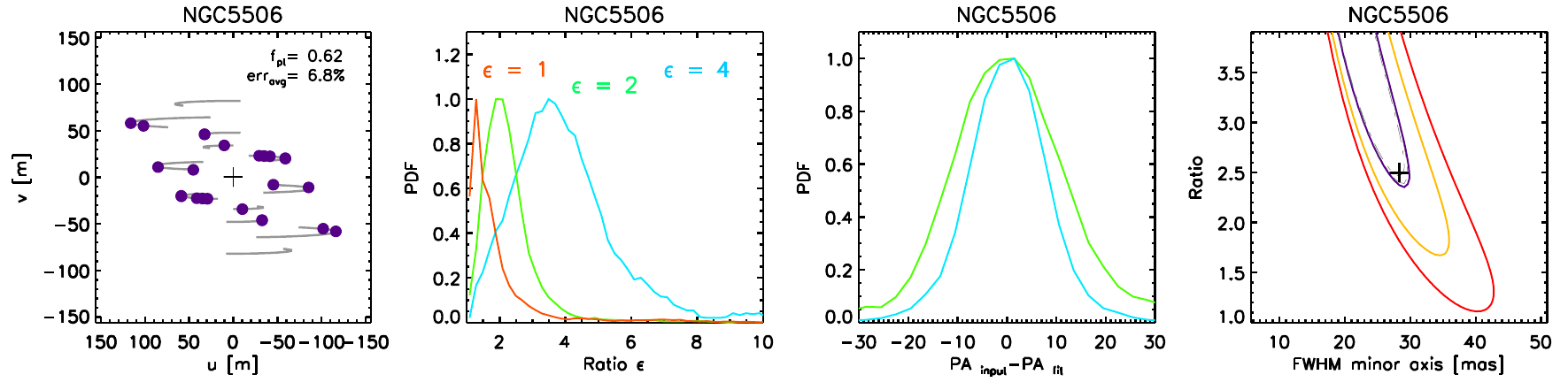}

	\caption{\label{fig:results1}: Results of our reliability analysis for sources where an elongation should be detectable given the observed $(u,v)$ coverage, signal/noise ratio and minimum visibility. 
	\quad {\it First column} ($u,v$) coverage of the object.
	The point source fraction ($f_{pt}$) and the typical uncertainty of the interferometric measurements ($err_{\rm avg}$ in percentage) are given for every object.\\
	{\it Second column}: Probability density function of the axis ratio for three different input axis ratios: $\epsilon=1$ in red, $\epsilon=2$ in green and $\epsilon=3$ in blue.
	For visualization purposes we have normalized the maximum value of the PDF's to a value of one.  \\
	{\it Third column}: Probability density function of the recovered position angle of the major axis minus the input position angle.
	The colors are the same as for the second column.
	Intentionally, we do not plot the resulting PDF for an input axis ratio equal to 1.\\
	{\it Fourth column}: Two dimensional confidence interval for the axis ratio and the FWHM of the minor axis from the fits using data from the actual observations. The 1-$\sigma$, 2-$\sigma$  and 3-$\sigma$ region are delimited by the purple, orange and red lines, respectively. 
	The gray dash-dotted lines denote contours inside the 1-$\sigma$ region.} 
\end{figure*}

\begin{figure*}
	\includegraphics[width=0.9\hsize]{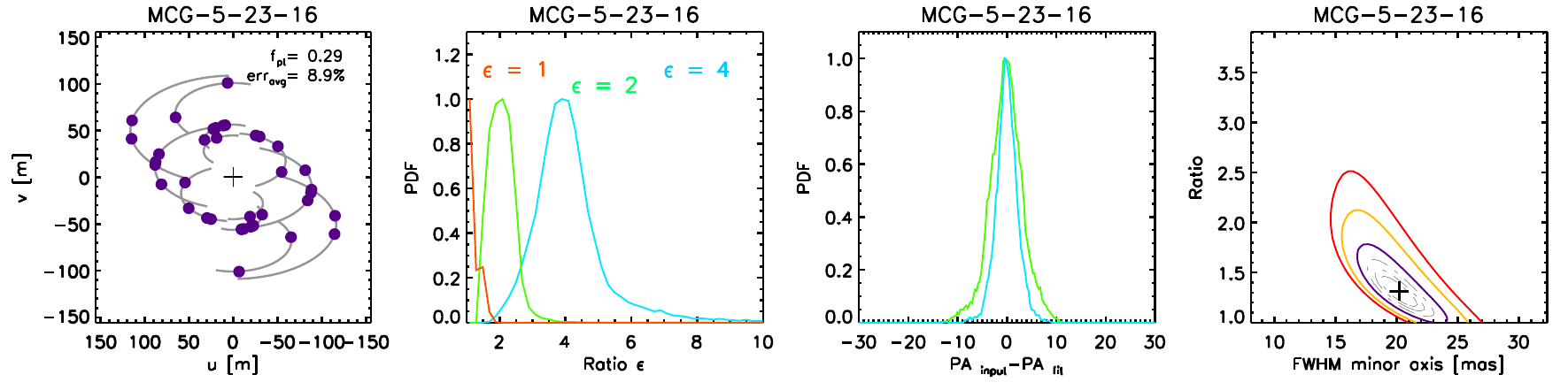}\\
	\includegraphics[width=0.9\hsize]{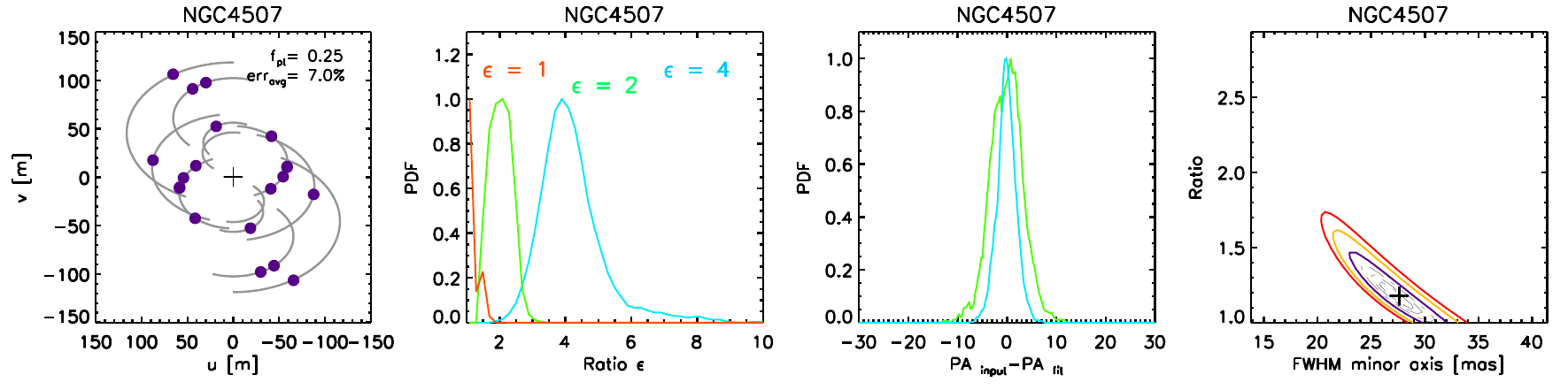}\\
	\caption{\label{fig:results2} Like Fig.~\ref{fig:results1}, but for sources where the mid-infrared emission is consistent with a nearly circular shape.}
\end{figure*}

\subsection{Reliability of model fits. The role of the ($u,v$) coverage}
\label{sec:reliability}

Using interferometric measurements to constrain a possibly elongated shape of the emission is not straightforward if the ($u,v$) coverage is very sparse and inhomogeneous.
Beam effects can easily dominate the shape of the reconstructed model image.

Therefore, we need to investigate the influence of the ($u,v$) coverage in the final determination of the axis ratio and the position angle.
For this, we set up a series of specific tests to assess the reliability of our fitting routines given the actual $(u,v)$ coverage of every object as well as the noise levels and the known minimum visibility (see Fig.~\ref{fig:flowchart} for an outline).

For this experiment, we first create models of elongated Gaussians, with a few specified axis ratios and all possible position angles (0-180$^{\circ}$ in 0.5$^{\circ}$ steps).
We set the major axis FWHM such that this component is marginally resolved (30 \%) at the shortest observed baselines.
We have also explored different size scales for the minor axis FWHM (15 \%, 60 \%) and the results regarding the reliability of the fit do not change significantly.
For this experiment, we include a point source to the model image with the known parameters.
This is important since elongated emission is much easier to detect if the source is well resolved than if the source is largely unresolved.

We then compute the Fourier transform of this model image and ``observe'' the model ten times by taking the visibilities at the same $(u,v)$ positions as the actual observations and adding Gaussian noise to the modeled data so that they are of the same signal/noise as the actual data. 
We apply our fitting routines on these mock observations to derive the best fit parameters and repeat this experiment for all sources.

As a last step we compare the best fit values of axis ratio and position angle with the known model input values and plot the probability of retrieving the correct axis ratio and input position angle for the different realizations of our model image. 
The ($u,v$) coverage, the probability density function (PDF) of the axis ratio and the PDF of the position angle from this experiment are shown for every source in Figs.~\ref{fig:results1},~\ref{fig:results2} and \ref{fig:results3} along with the $\chi^2$ plane of the fit to the actual data.

These figures allow us to judge for each source how significant the result of our model fit is.
As a (somewhat arbitrary) distinction between the reliable (Fig.~\ref{fig:results1} and Fig.~\ref{fig:results2}) and less reliable (Fig.~\ref{fig:results3}) objects we use as a criterion whether the data would allow us to discriminate between a round source ($\epsilon=1$) and an elongated source with an axis ratio of 2. 
This can be easily read off from the center-left plot for each source: if the PDFs for the round ($\epsilon=1$) and the $\epsilon=2$ model source do not overlap at 1 $\sigma$ (68.3 \%), we consider this set of observations as a reliable means of detecting elongated emission.

The results from this experiment can be summarize as follows:

1) We cannot recover reliable axis ratios or position angles for objects with an ($u,v$) coverage that consists of measurements along essentially two directions (e.g., ESO 323-77, NGC5995, see Fig.~\ref{fig:results3}) or if the extended emission is overresolved (e.g, NGC 5128, see Fig.~\ref{fig:results3}).

2) It is also not possible to determine the axis ratio or position angle for objects with a point source fraction above 70 \% and with typical uncertainties of about 10 \% (e.g, NGC 3218, see Fig.~\ref{fig:results3}).

3) The position angle can be determined well for the objects shown in Fig.~\ref{fig:results1} and Fig.~\ref{fig:results2}, in these objects the ($u,v$) coverage has precise measurements along  three or more well spaced directions in the ($u,v$) plane.
We observe from their respective PDFs that for any direction, the position angle can be recovered with an accuracy of about 10 degrees.

Note that we only need to test the reproducibility of the axis ratio test to decide whether we can recover elongated emission since all sets of observations with well recoverable axis ratios also have well defined position angles while the inverse is not true.

\begin{figure*}
	\includegraphics[width=0.9\hsize]{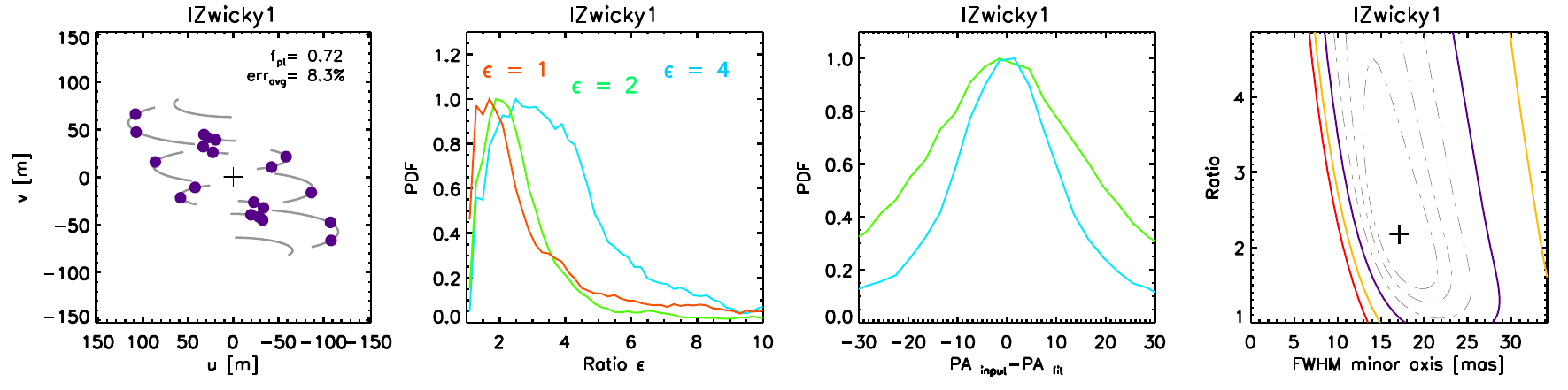}\\
	\includegraphics[width=0.9\hsize]{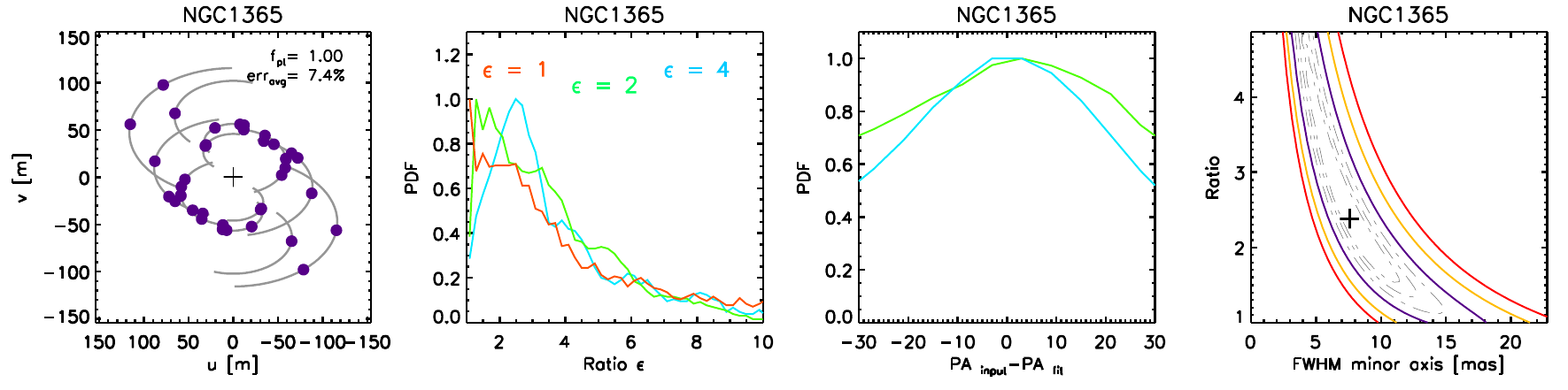}\\
	\includegraphics[width=0.9\hsize]{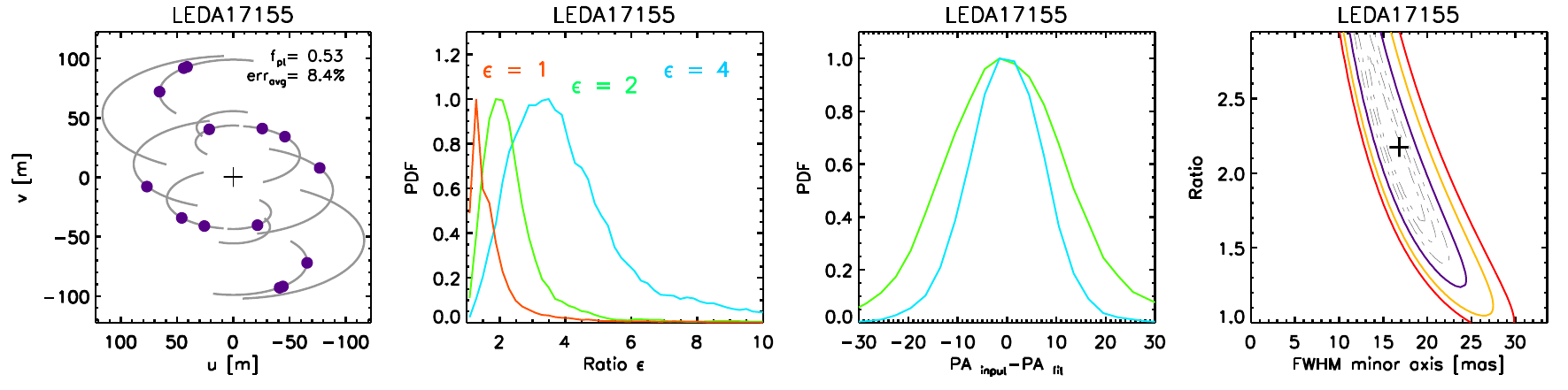}\\
	\includegraphics[width=0.9\hsize]{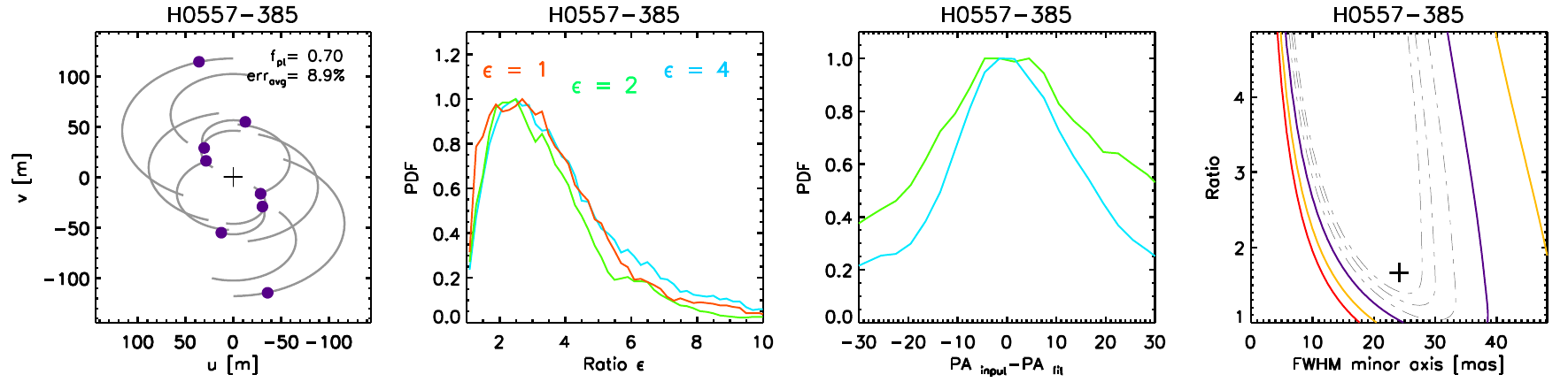}\\
	\includegraphics[width=0.9\hsize]{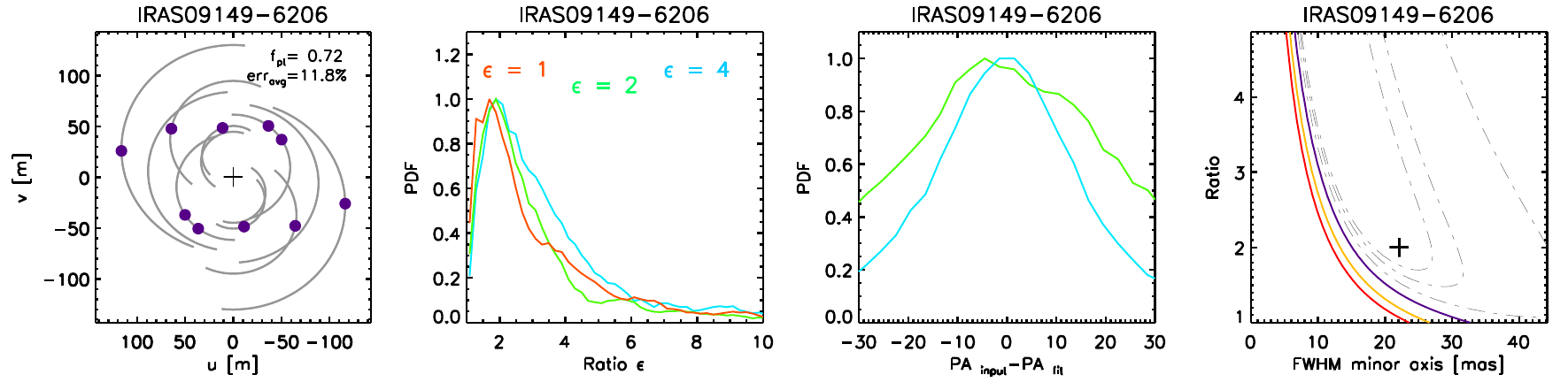}
	\caption{\label{fig:results3} Like Fig.~\ref{fig:results1}, but for sources where an elongation is not reliably detectable due to either bad $(u,v)$ coverage, low signal/noise or high minimum visibility (or a combination of these factors).}
\end{figure*}

\begin{figure*}
	\ContinuedFloat
	\includegraphics[width=0.9\hsize]{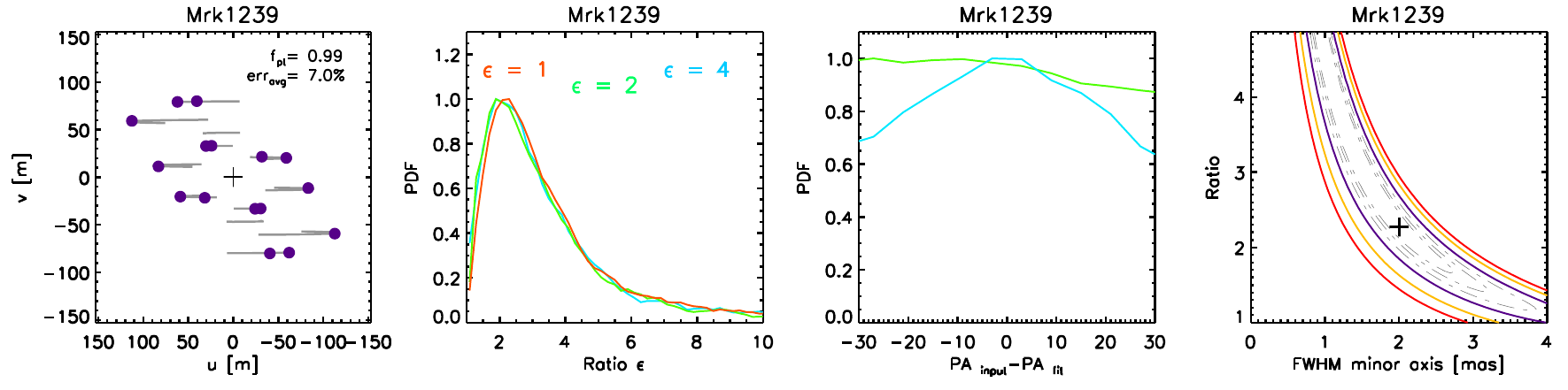}\\
  	\includegraphics[width=0.9\hsize]{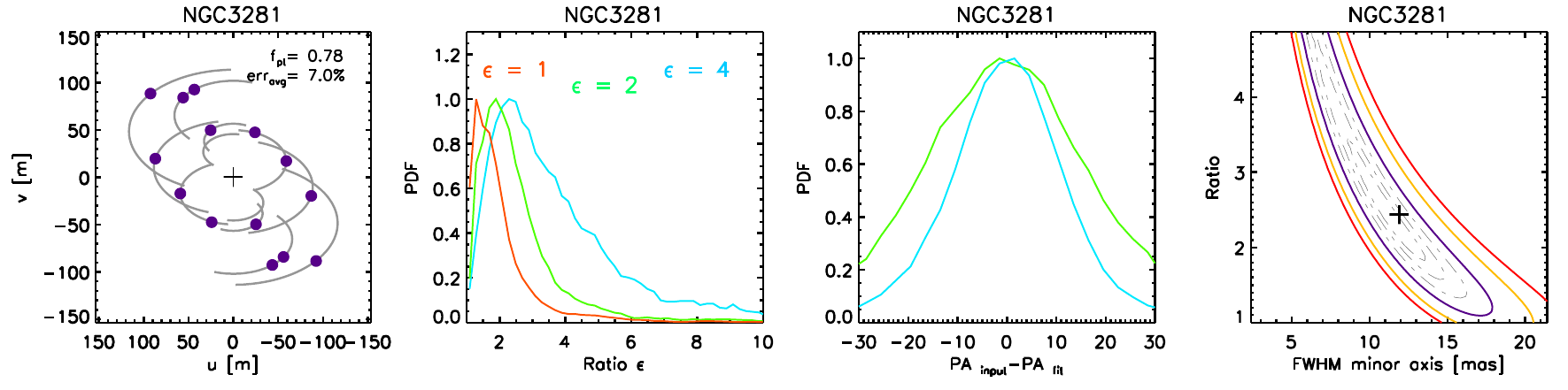}\\
	\includegraphics[width=0.9\hsize]{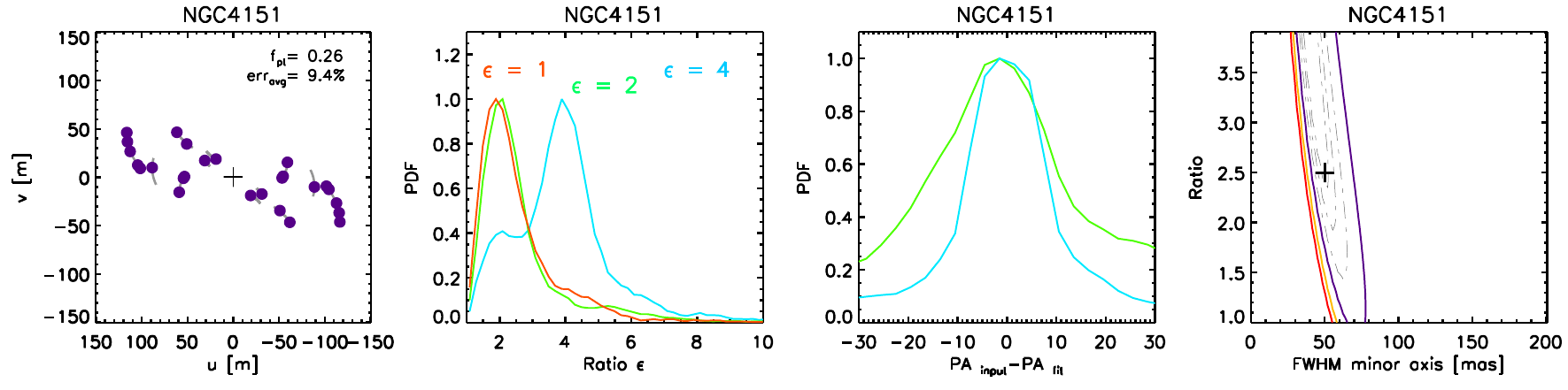}\\
	\includegraphics[width=0.9\hsize]{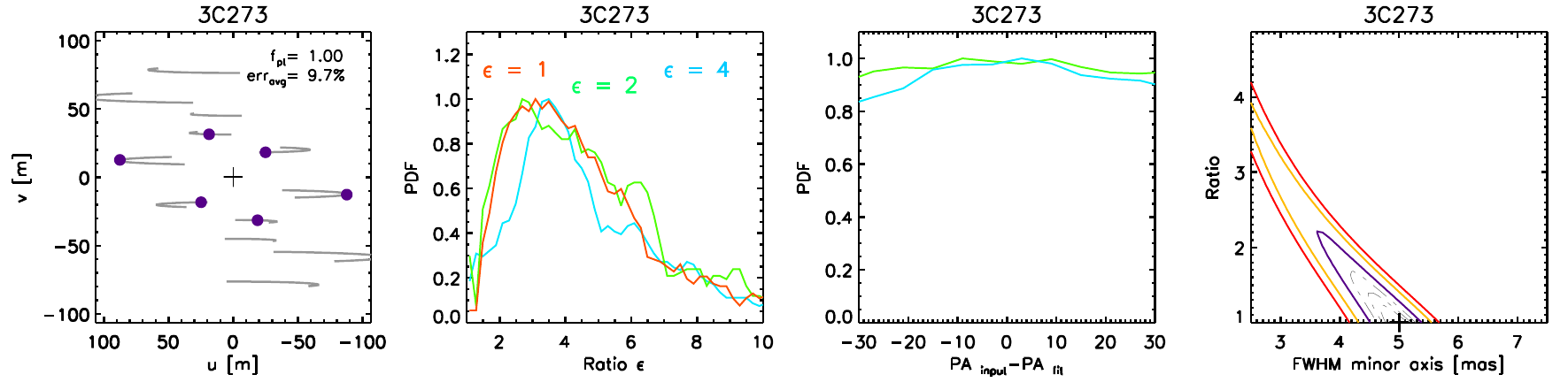}\\
	\includegraphics[width=0.9\hsize]{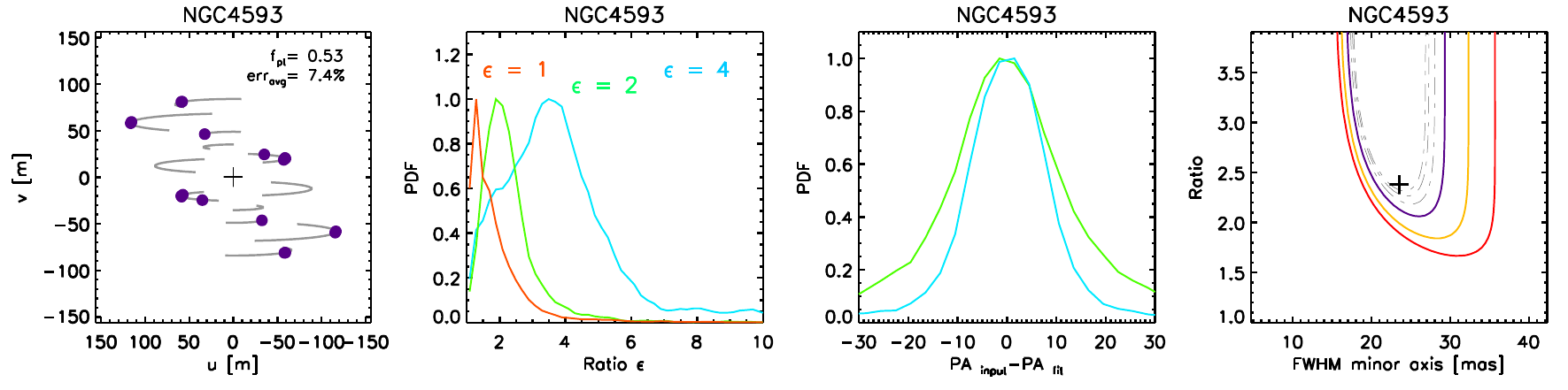}
	\includegraphics[width=0.9\hsize]{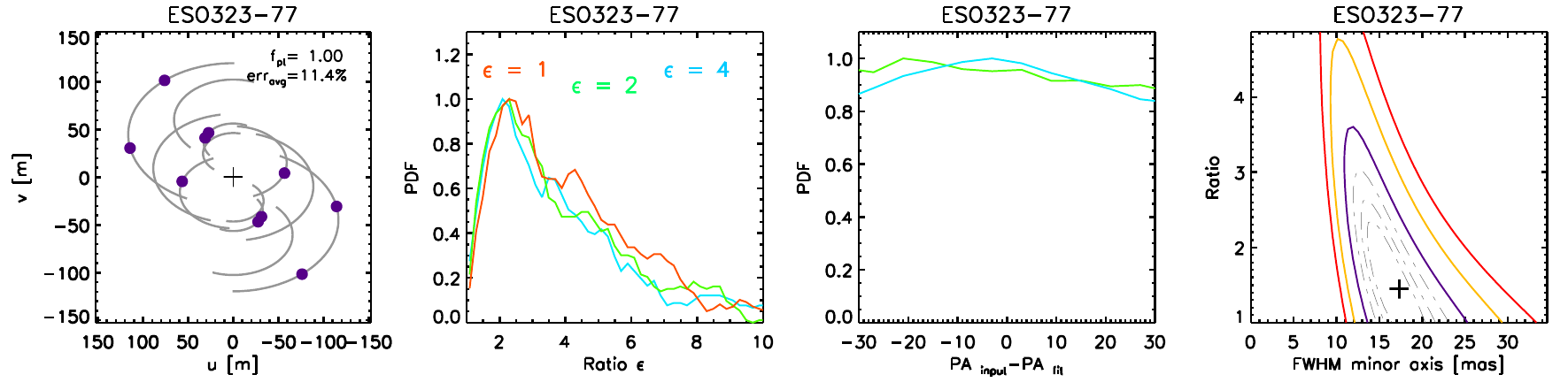}\\
	\caption{ --- {\em continued}}
\end{figure*}

\begin{figure*}
	\ContinuedFloat
	\includegraphics[width=0.9\hsize]{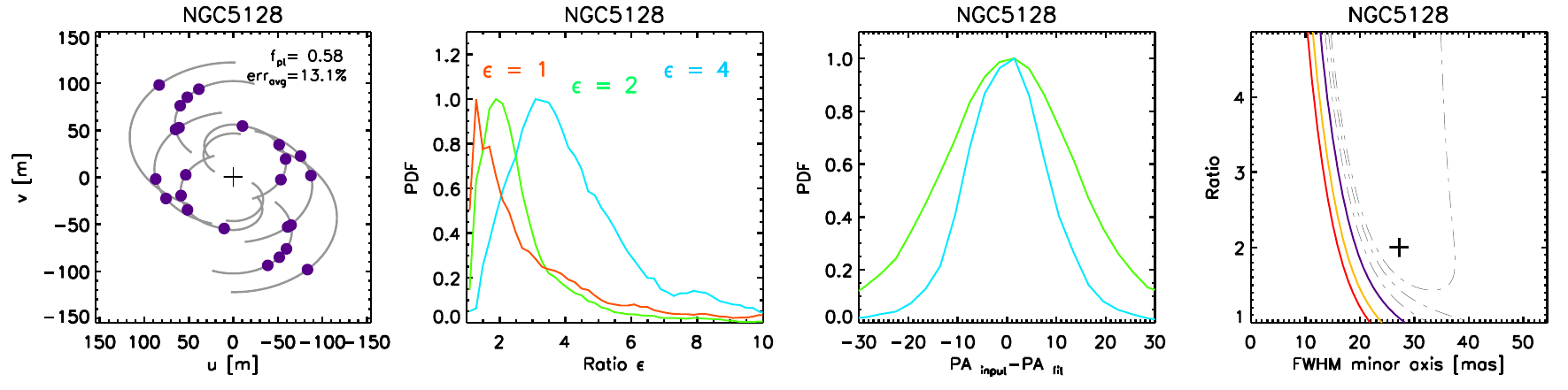}\\
	\includegraphics[width=0.9\hsize]{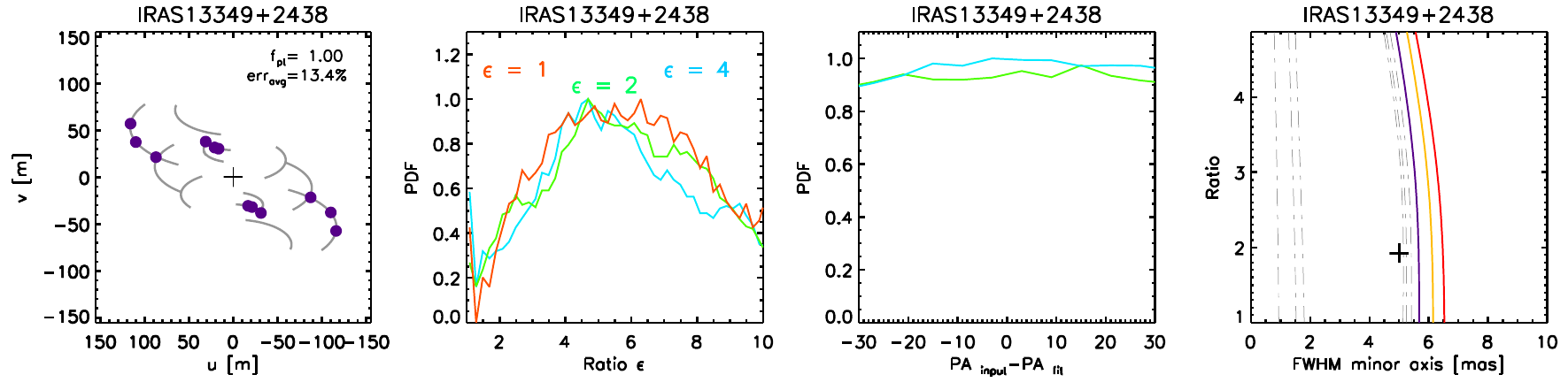}\\
	\includegraphics[width=0.9\hsize]{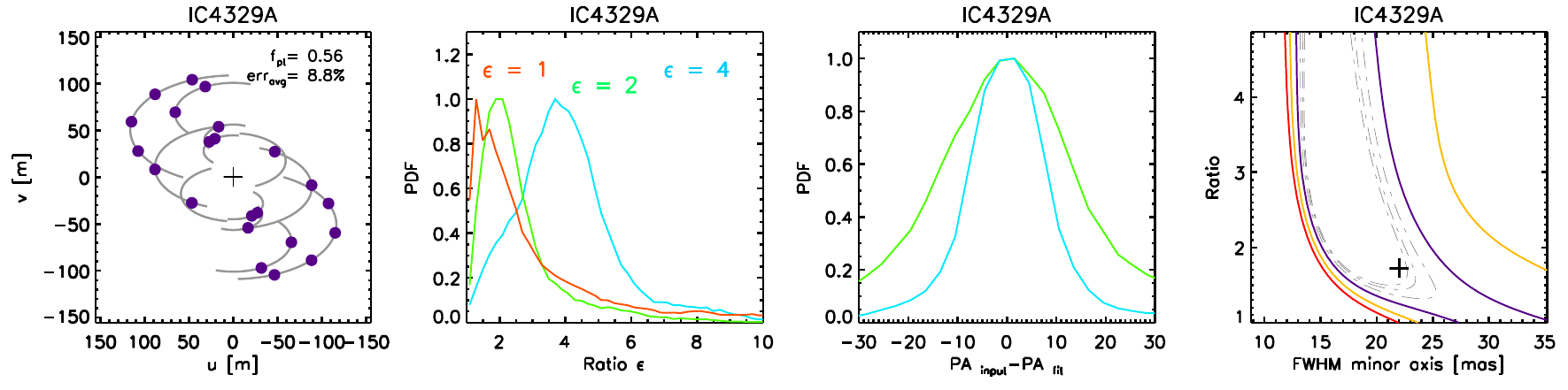}\\
	\includegraphics[width=0.9\hsize]{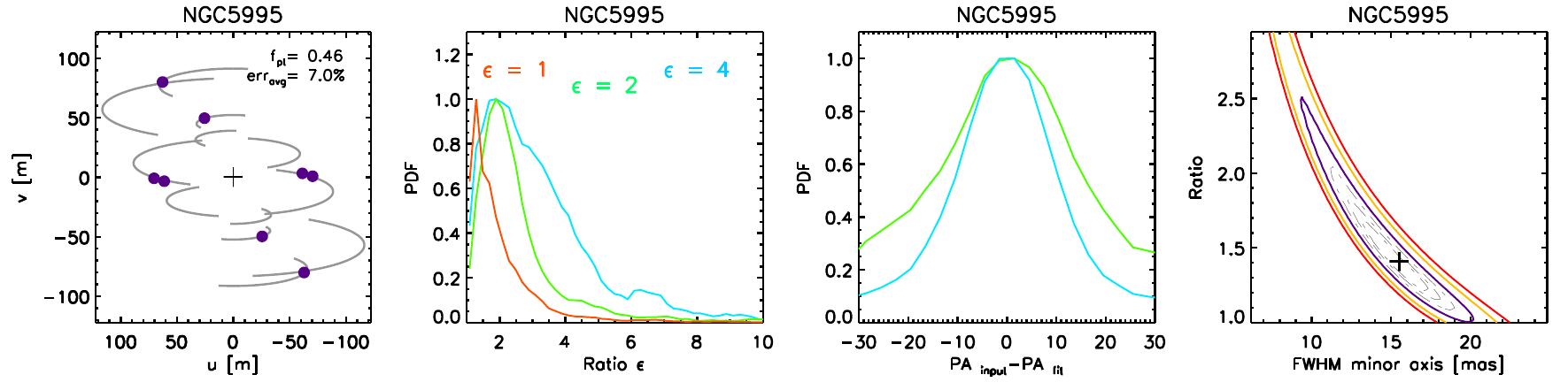}\\
	\includegraphics[width=0.9\hsize]{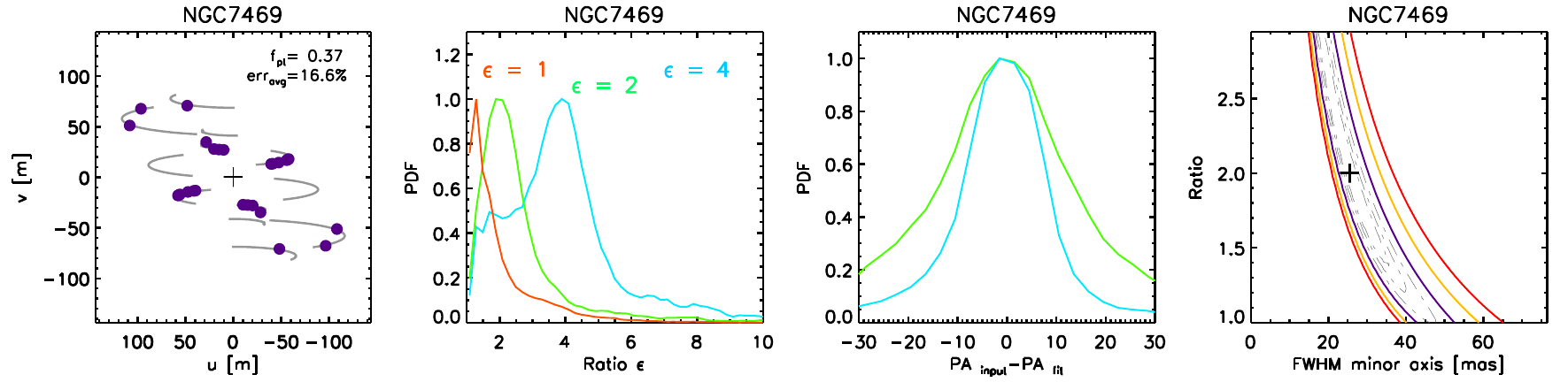}
	\caption{ --- {\em continued}}
\end{figure*}

\begin{table*}
\centering
\small
\begin{tabular}{ c | c c c c c | c | c c c }
\hline
\hline
\multicolumn{10}{c}{\multirow{2}*{ELONGATED}} \T \B\\
\multicolumn{10}{c}{}  \B \\
\hline
 & \multicolumn{5}{c |}{Parameters of best fit model} &  & \multicolumn{3}{c}{PA}  \T \\
Source & $F_{pt}$ & minor axis $\theta_m$ & PA & Ratio & $F_{gs}$ & $\chi_{red}^2$  & Polar axis & NLR & Radio jet \\
 & [Jy] & [mas] & [Degree] & $\varepsilon=\Theta_M/\theta_m$ & [Jy] & & [Degree] & [Degree] & [Degree] \B \\
\hline
\hline
     NGC424 &  0.30 $\pm$ 0.01 &  15.9 $_{-2.6 }^{+3.9 }$ $(  3.4_{-0.6}^{+0.8})$ & 157.4 $\pm$ 5.4  &   1.9$_{ -0.5}^{+  0.6}$ &  0.35 $\pm$ 0.01 &  3.13 & 133 $^{(a)}$ & - & unresolved $^{(e)}$ \T \B \\
    NGC1068 &  5.23 $\pm$ 0.19 &  71.2 $_{-6.0 }^{+10.4}$ $(  5.0_{-0.4}^{+0.7})$ & 155.8 $\pm$ 2.9  &   2.3$_{ -0.5}^{+  0.5}$ & 10.81 $\pm$ 0.19 &  6.53 & 5 $^{(b)}$   & 30 $^{(c)}$ & 0 $^{(d)}$  \T \B \\
    NGC3783 &  0.19 $\pm$ 0.01 &  15.7 $_{-1.3 }^{+1.3 }$ $(  3.2_{-0.3}^{+0.3})$ & 119.9 $\pm$ 3.6  &   3.1$_{ -0.6}^{+  0.8}$ &  0.49 $\pm$ 0.01 &  3.5 & 135 $^{(b)}$ & 160 $^{(c)}$ & unresolved $^{(d,g)}$\T \B \\
   Circinus &  1.01 $\pm$ 0.11 &  47.6 $_{-15.0}^{+13.1}$ $(  1.0_{-0.3}^{+0.2})$ & 106.6 $\pm$ 2.4  &   1.9$_{ -0.4}^{+  1.1}$ & 9.08 $\pm$ 0.11 &  2.41 & 135 $^{(b)}$ & 128 $^{(c)}$ & 115 $^{(f)}$  \T \B \\
    NGC5506 &  0.57 $\pm$ 0.01 &  28.6 $_{-9.4 }^{+3.9 }$ $(  4.0_{-1.3}^{+0.5})$ &   8.3 $\pm$ 9.0  &   2.5$_{ -0.4}^{      }$ &  0.47 $\pm$ 0.01 &  3.18 & 163 $^{(b)}$ & 22 $^{(c)}$ & 155 $^{(h)}$ \T \B \\
\hline
\hline
\multicolumn{9}{c}{\multirow{2}*{NEARLY CIRCULAR}} \T \B\\
\multicolumn{9}{c}{}  \B \\
\hline
\hline
MCG-5-23-16 &  0.20 $\pm$ 0.01 &  20.2 $_{-3.3 }^{+4.2 }$ $(  3.8_{-0.6}^{+0.8})$ &  88.5 &   1.3$_{ }^{+  0.5}$ &   0.39 $\pm$ 0.01 &  1.48 & -  & - & 169 $^{(i)}$ \T \B \\
    NGC4507 &  0.16 $\pm$ 0.01 &  27.6 $_{-5.2 }^{+4.6 }$ $(  6.9_{-1.3}^{+1.1})$ &  97.7 &   1.2$_{ }^{+  0.3}$ &   0.48 $\pm$ 0.01 &  7.2 & 127 $^{(b)}$   & 143 $^{(c)}$  & unresolved $^{(j)}$ \T \B \\
\hline
\hline
\multicolumn{9}{c}{\multirow{2}*{UNDETERMINED}} \T \B\\
\multicolumn{9}{c}{}  \B \\
\hline
\hline
\multicolumn{2}{c}{IZwicky1}           & \multicolumn{2}{c}{NGC1365 (u)}  & \multicolumn{3}{c}{IRAS05189-2524} & \multicolumn{3}{c}{H0557-385} \T\\
\multicolumn{2}{c}{IRAS09149-6206}     & \multicolumn{2}{c}{Mrk1239 (u)}  & \multicolumn{3}{c}{NGC3281}        & \multicolumn{3}{c}{NGC4151}   \T\\
\multicolumn{2}{c}{3C273 (u)}          & \multicolumn{2}{c}{NGC4593}      & \multicolumn{3}{c}{ESO323-77 (u)}  & \multicolumn{3}{c}{NGC5128}   \T\\
\multicolumn{2}{c}{IRAS13349+2438 (u)} & \multicolumn{2}{c}{IC4329A}      & \multicolumn{3}{c}{NGC5995}        & \multicolumn{3}{c}{NGC7469}   \T\\
\hline
\hline
\end{tabular}
 \caption{Parameters of the best fit models. 
 \small 
 {\it Source:} name.
 {\it $F_{pt}$:} Flux of the point source.
 {\it minor axis $\theta_m$:} Full Width at Half Maximum of the minor axis.
 {\it PA:} Position angle of the major axis of the Gaussian.
 {\it Ratio ($\epsilon$):} Ratio between the major and minor axis. 
 To avoid degeneracies, we only take $\epsilon\geq1$. 
 Axis ratios with $\epsilon < 1$ produce the same elongation as $1/\epsilon$ but with PA rotated by $90^\circ$. 
 {\it $F_{gs}$:} Flux of the elongated Gaussian component.
 {\it $\chi^2_{reduced}$:} Reduced chi-square value.
 {\it PA Polar axis:} Position angle of the polar axis inferred from optical polarimetry. 
 For Type 2 objects we add 90 degrees to the true position angle given by optical polarization (see text).
 {\it PA NLR:} Position angle of the symmetry axis of the ionization cone.
 {\it PA Radio jet}: Position angle of the radio emission from the jet. 
 Objects with marginally resolved or unresolved 12 $\mu$m emission are marked with (u).
 {\bf References:} a)\citet{2012ApJ...755..149H}; b) \citet{2014MNRAS.441..551M}; c) \citet{2013ApJS..209....1F}; d) \citet{2000ApJ...537..152K}, e) \citet{2000ApJ...529..816M}; f) \citet{1998MNRAS.297.1202E}; g) \citet{2010MNRAS.401.2599O}, h) \citet{2010MNRAS.404.1966X}; i) \citet{2009ApJ...703..802M}; j) \citet{1998ApJ...497..133B}    }
 \label{table:Tparelong}
\end{table*}
\normalsize

\section{Results and discussion}

\begin{figure}
\centering
\begin{minipage}{0.5\hsize}
  \centering
    \includegraphics[width=\hsize]{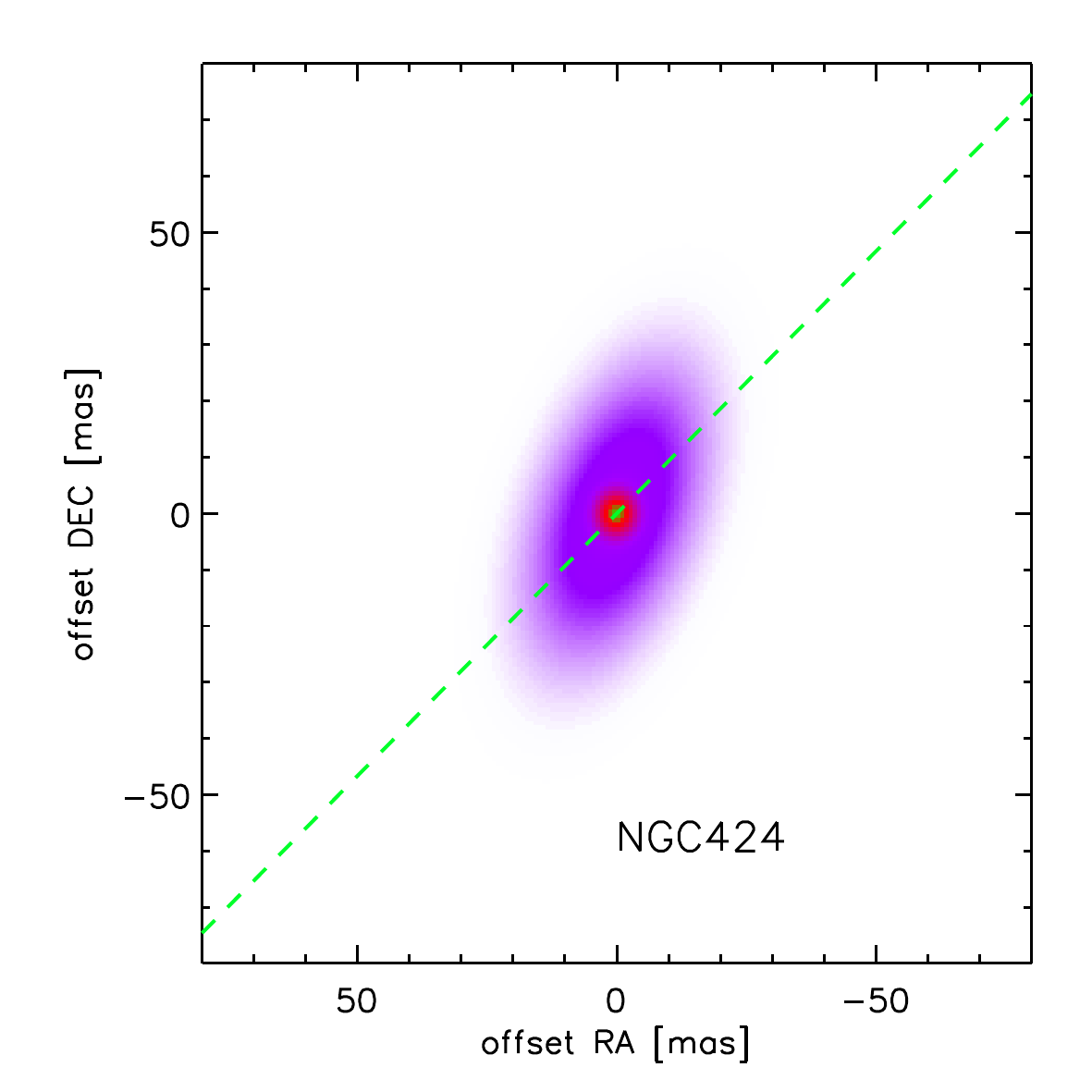}
   \includegraphics[width=\hsize]{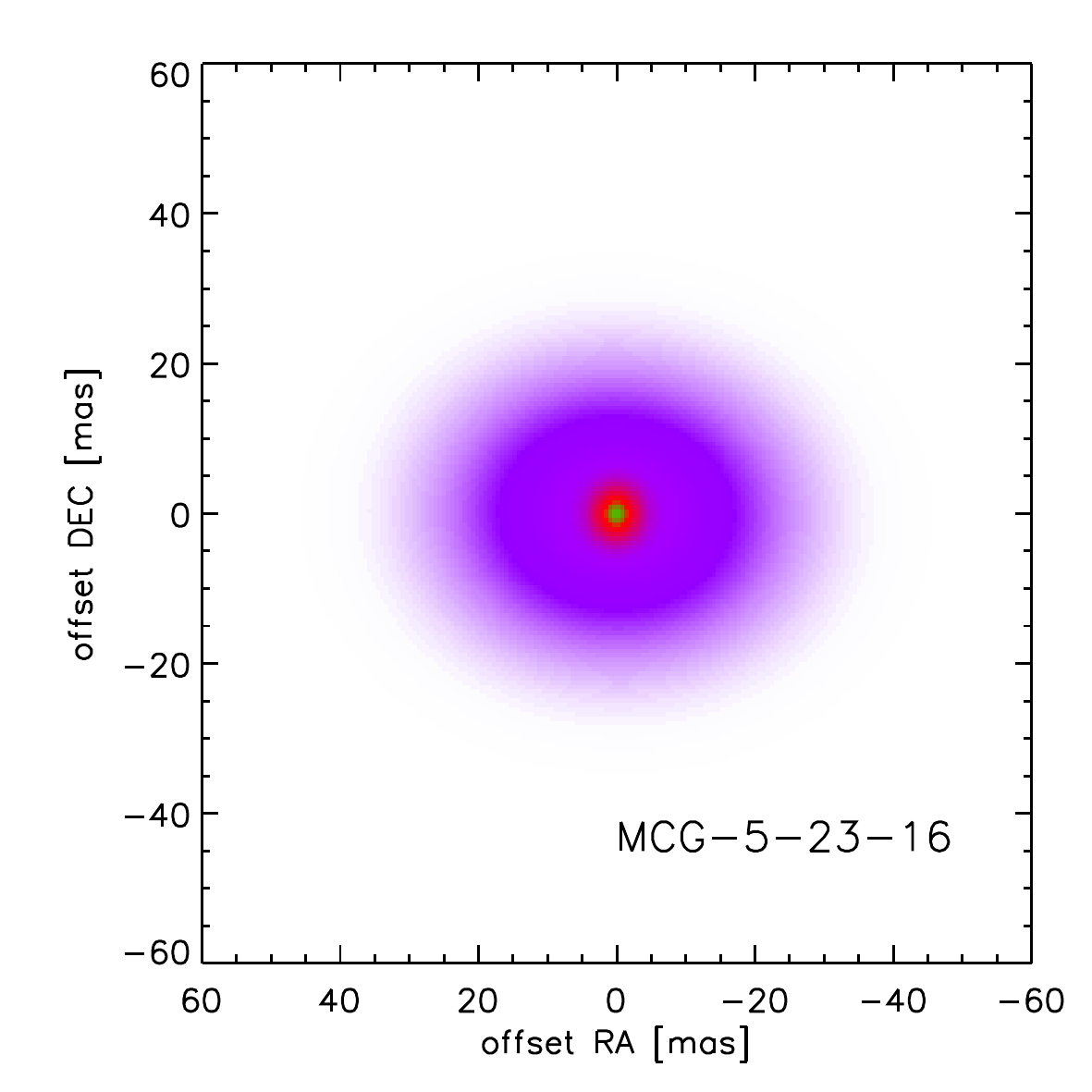}
    \includegraphics[width=\hsize]{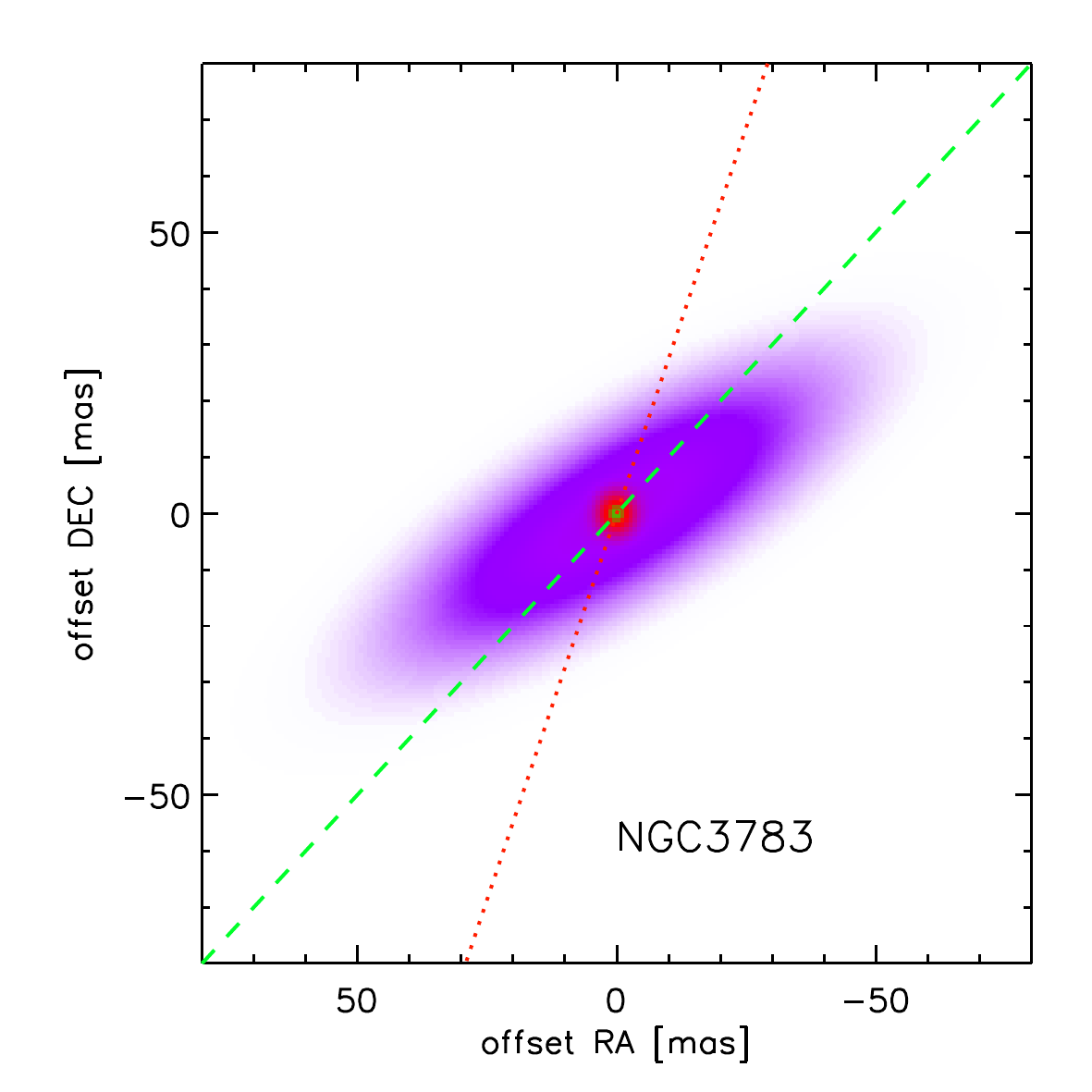}
    \includegraphics[width=\hsize]{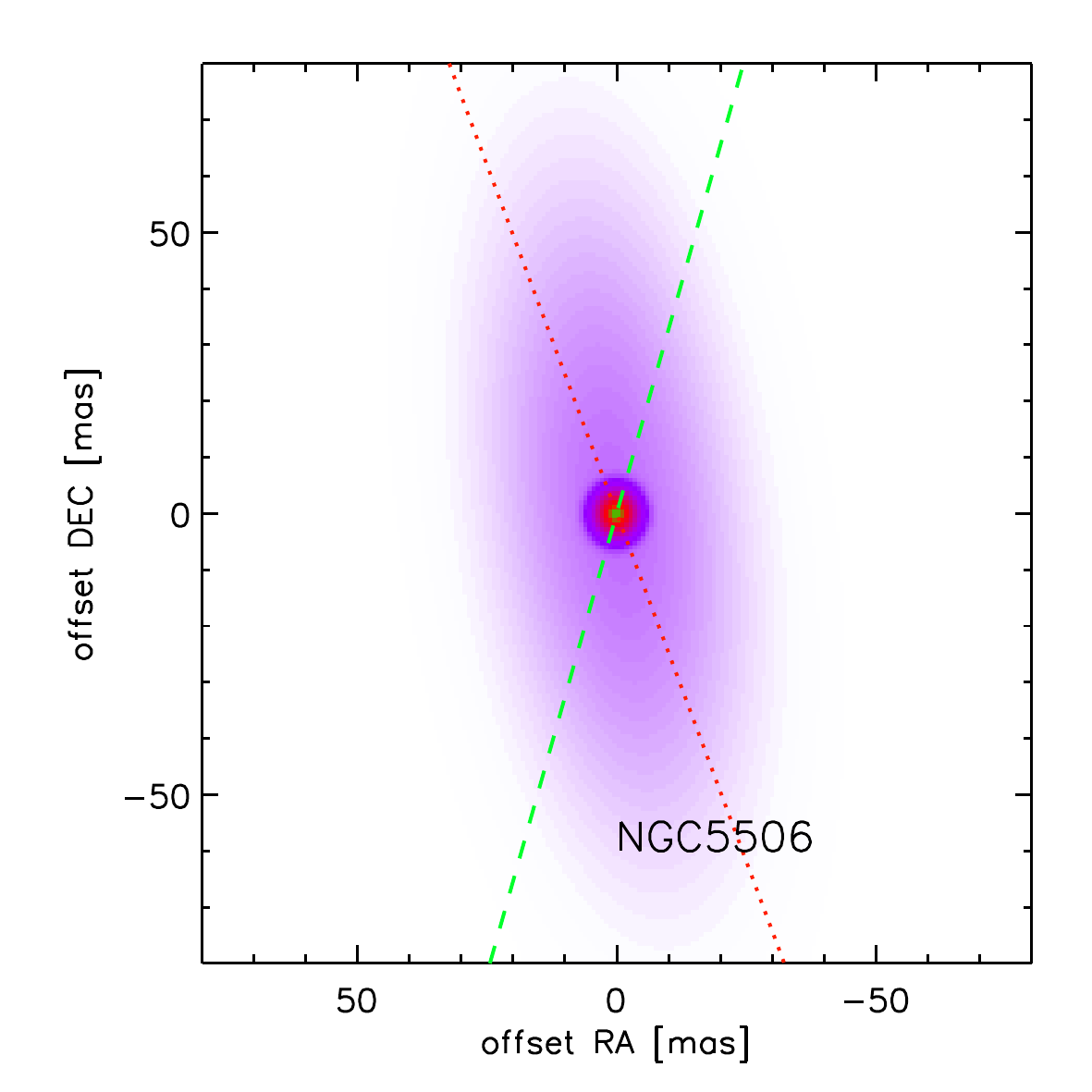}
\end{minipage}%
\begin{minipage}{0.5\hsize}
  \centering
    \includegraphics[width=\hsize]{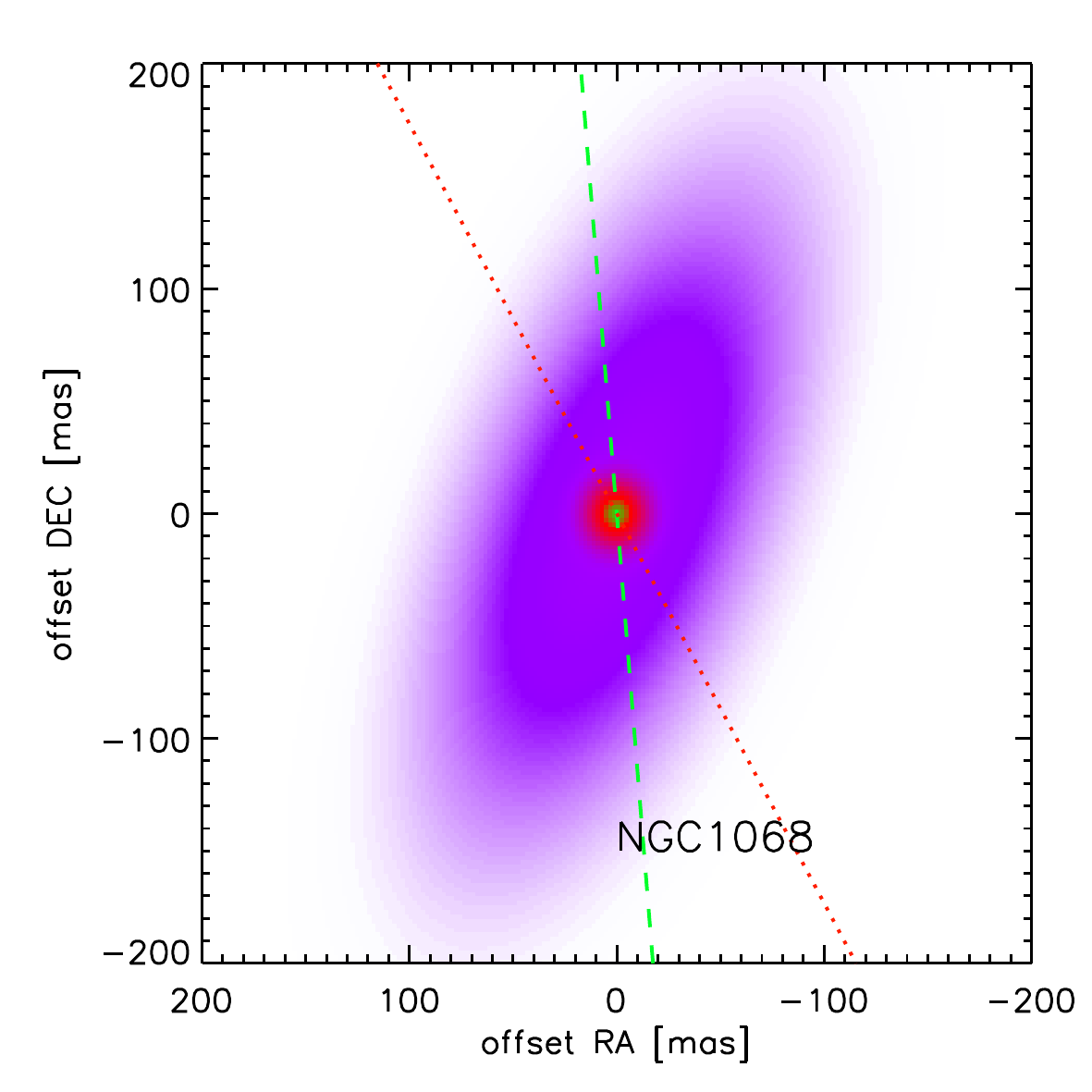}
        \includegraphics[width=\hsize]{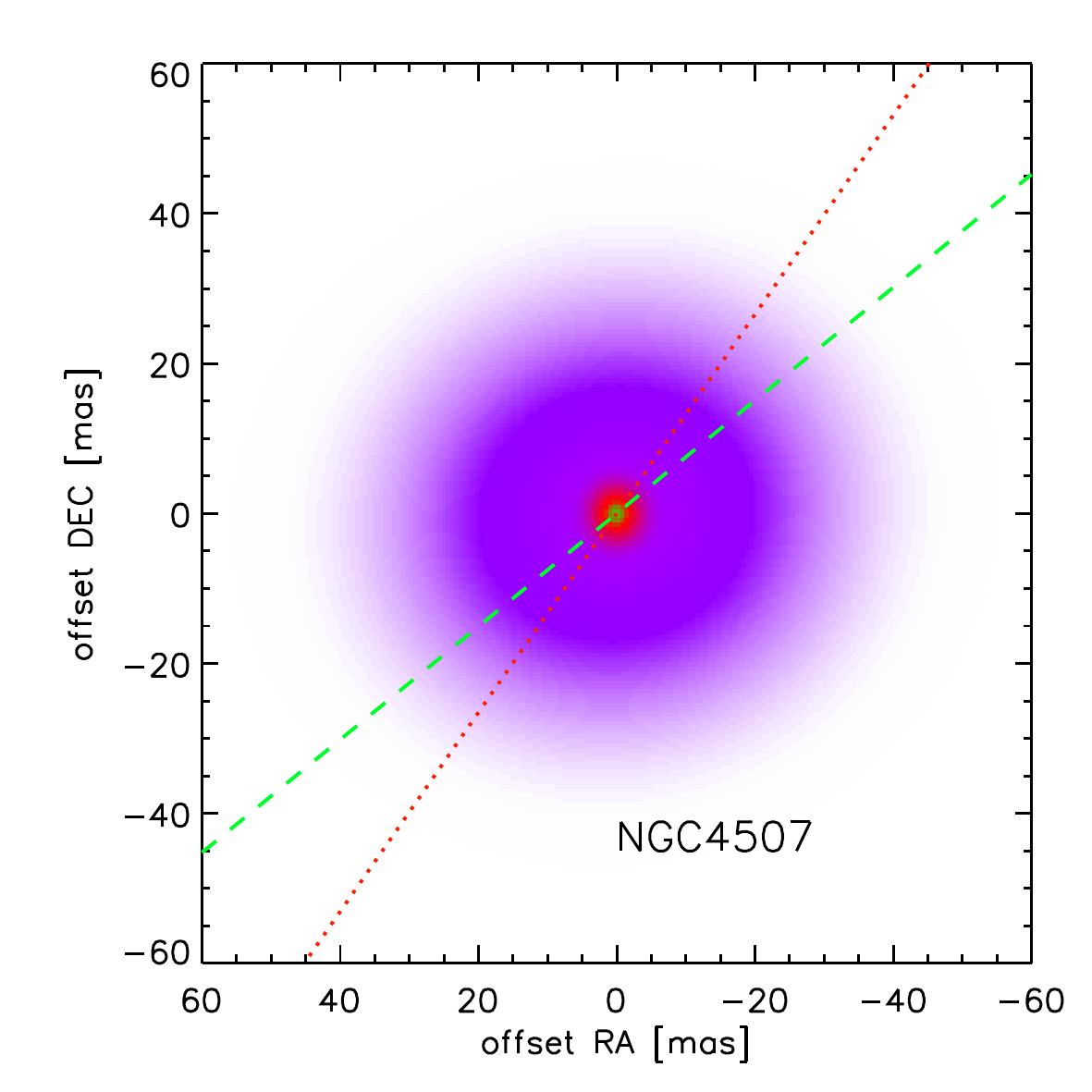}
    \includegraphics[width=\hsize]{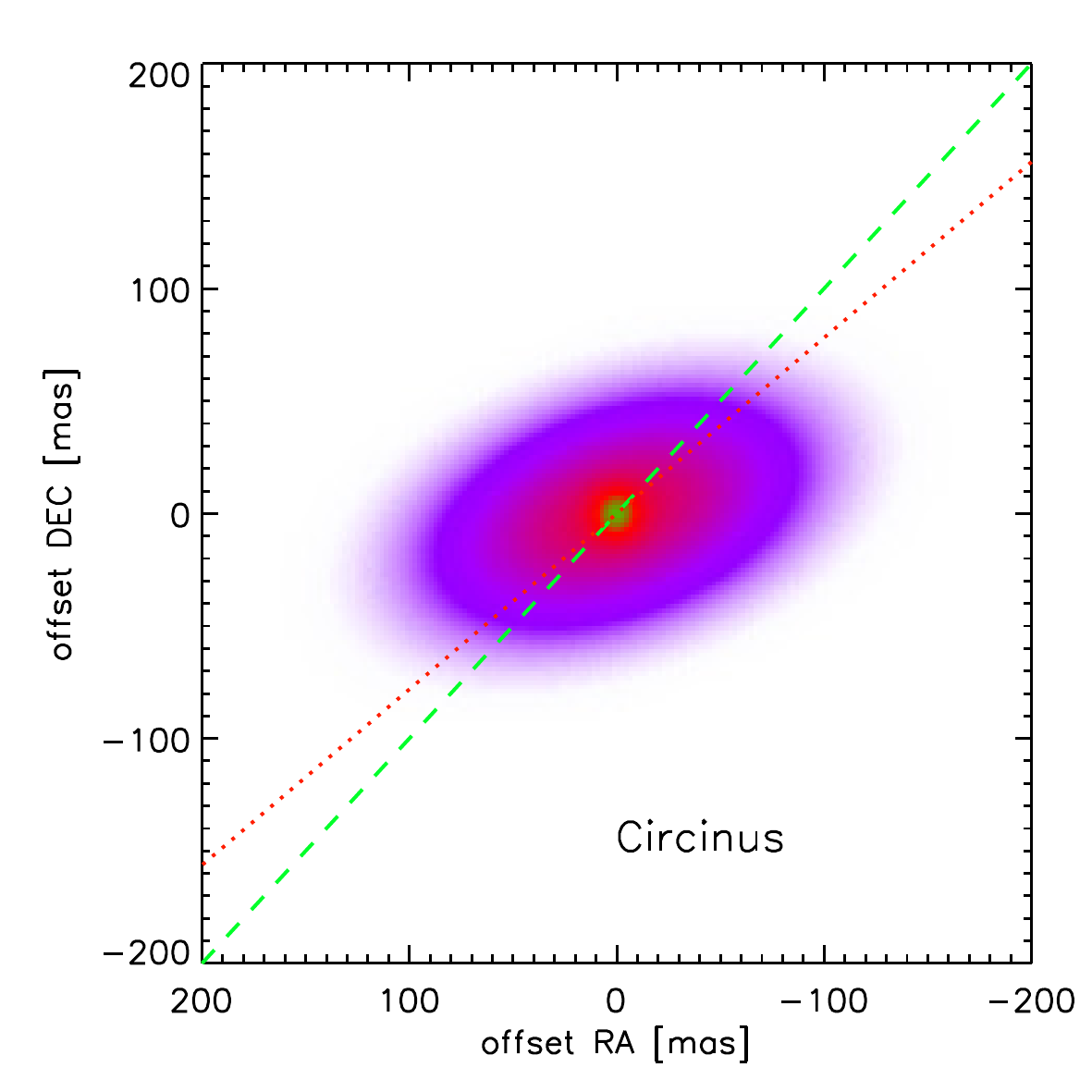}
\end{minipage}%
 \captionof{figure}{Best fit models for the elongated objects.
 For every object we show the 12 $\mu$m image obtained from our best fit model using a squared root scale. We add as a reference for the polar axis  of the system, when available, the PA obtained from optical polarimetry (green dashed line) and from the symmetry axis of the ionization cone (red dotted line).}
  \label{fig:elong}
\end{figure}


Based on our reliability analysis (Sec.~\ref{sec:reliability}) and Figs.~\ref{fig:results1} and \ref{fig:results2} we find seven sources where the existing observations would allow us to reliably detect elongated emission.
For these sources, we give the best fitting parameters for the elongated Gaussian model fits together with auxiliary data in Tab.~\ref{table:Tparelong}.

Elongated mid-IR emission on the parsec-scale was found previously in four of the sources in this sample: NGC~424 \citep{2012ApJ...755..149H}, NGC~3783 \citep{2013ApJ...771...87H}, NGC~1068 \citep{2014A&A...565A..71L} and the Circinus~galaxy \citep{2014A&A...563A..82T}.
These are among the brightest and very well studied sources ($\gtrsim$ 30 $(u,v)$ points) where elongation and position angle can be constrained well.
Our results (Tab.~\ref{table:Tparelong}) are in excellent agreement with the published ones, confirming the reliability of the detected elongations.
For NGC1068 and Circinus higher spatial resolution data show a distinct disk-like structure which contributes  $\sim 20$\% to the total flux.
This disk has a different PA than the large scale structure we model here.
We ignore the small-scale structure for our large scale fits since we would not be able to resolve such a structure in any of the other sources due to the lower spatial resolutions achieved here.
The FWHM sizes and PAs derived from our large-scale analysis only, are however, very similar to the ones derived with the more complex disk + large scale model.

Our reliability analysis has shown that we should also be able to detect elongated emission in MCG~-05-23-16, NGC~4507 and NGC~5506.
In NGC~5506 our fits indicate that the emission is indeed elongated.
The confidence limit of the fit show that the axis ratio is inconsistent with 1 (circular) at 3$\sigma$ level (Fig~\ref{fig:results1}).
The best fit elongated model has an axis ratio of 2.5 with a well defined lower limit but with its major axis less well determined. 
The position angle of the major axis of the elongation is 8.3$^\circ\pm 9$.

For MCG-5-23-16 and NGC4507, on the other hand, the 12 $\mu$m emission at subparsec scales is consistent with a near circular shape, see Fig.~\ref{fig:results2}.
The best fit axis ratios are 1.3 $^{+0.5}$ and 1.2$^{+0.3}$ for the two sources respectively.

For the remaining 16 objects the ($u,v$) coverage and the typical uncertainties do not allow us to recover the shape of the input brightness distribution.

\subsection{Polar or equatorial emission?}

Determining the axis ratio and position angle of the mid-infrared emission provides valuable information about the geometrical shape of the dusty emission, but in order to learn more about its nature  we need to locate the position angle of the dust emission with respect to a known axis of the system.  
[OIII] line emission produced along the ionization cone can in principle be used to determine the direction of the polar axis. 
\citet{2013ApJS..209....1F} were able to identify the direction of the ionization cones by using bi-conical outflow models to describe the kinematic information of the NLR, given by long-slit spectra from the Space Telescope Imaging Spectrograph and [OIII] imaging.
We can in principle assume that the symmetry axis of such a cone is a reasonable estimate of the polar axis of the system. 
The values reported for our sources by \citet{2013ApJS..209....1F}  are shown in Table~\ref{table:Tparelong}.
Since the axis of the ionization cone might be considered as model dependent\footnote{In the case of NGC~3783, the model reported by \citet{2013ApJS..209....1F} differs in inclination from the model provided by \citep{2011ApJ...739...69M}, although the position angles are rather similar (-20$^\circ$ and -177$^\circ$ respectively).}, we additionally obtain an estimate of the system axis from optical polarimetry. 
As discussed by \citet{2014MNRAS.441..551M}, the degree of polarization observed in Type 1 sources is relatively small, but any position angle detected for the optical polarization should be roughly parallel to the polar axis of the system. 
For Type 2 sources the degree of polarization should be higher and with a position angle of the polarized light roughly perpendicular to the polar axis \citep[e.g.][]{1999ApJ...518..676K}.
This explanation could be a simplified version of reality since there might also be a transition region between type 1 and type 2 objects where the position angle follows a different behavior. 
Keeping in mind this possible ambiguity, we nevertheless use also the polarization measurements to infer a system axis and list them in column 8 of Table 1 and plot them in Fig. 5 as green dashed lines. 
We find a reasonable agreement between the system axis determined by the NLR modelling and polarimetry with the largest discrepancy of $39^{\circ}$ between the two inferred angles in the case of NGC5506. 
For our purpose of determining whether the MIR dust emission is rather polar or equatorial elongated, this is sufficient accuracy.

In principle, we could also obtain information about the polar system axis from observing the direction of an existing jet. 
Synchrotron emission from the jet has been observed intensively at radio bands.
We have collected information about the position angle of the jet from the literature and compare them with the values obtained for the polar axis from optical polarimetry and from the NLR (see Table~\ref{table:Tparelong}).
For the objects NGC4507, NGC3783 and NGC424, the emission is unresolved and therefore we cannot obtain any relevant information about the PA. 
In the case of Circinus, NGC1068 and NGC5506, the position angle of the jet agrees very well with the values obtained from optical polarimetry and a bit less with the values from the NLR.

\begin{figure}
   \centering
   \includegraphics[width=\hsize]{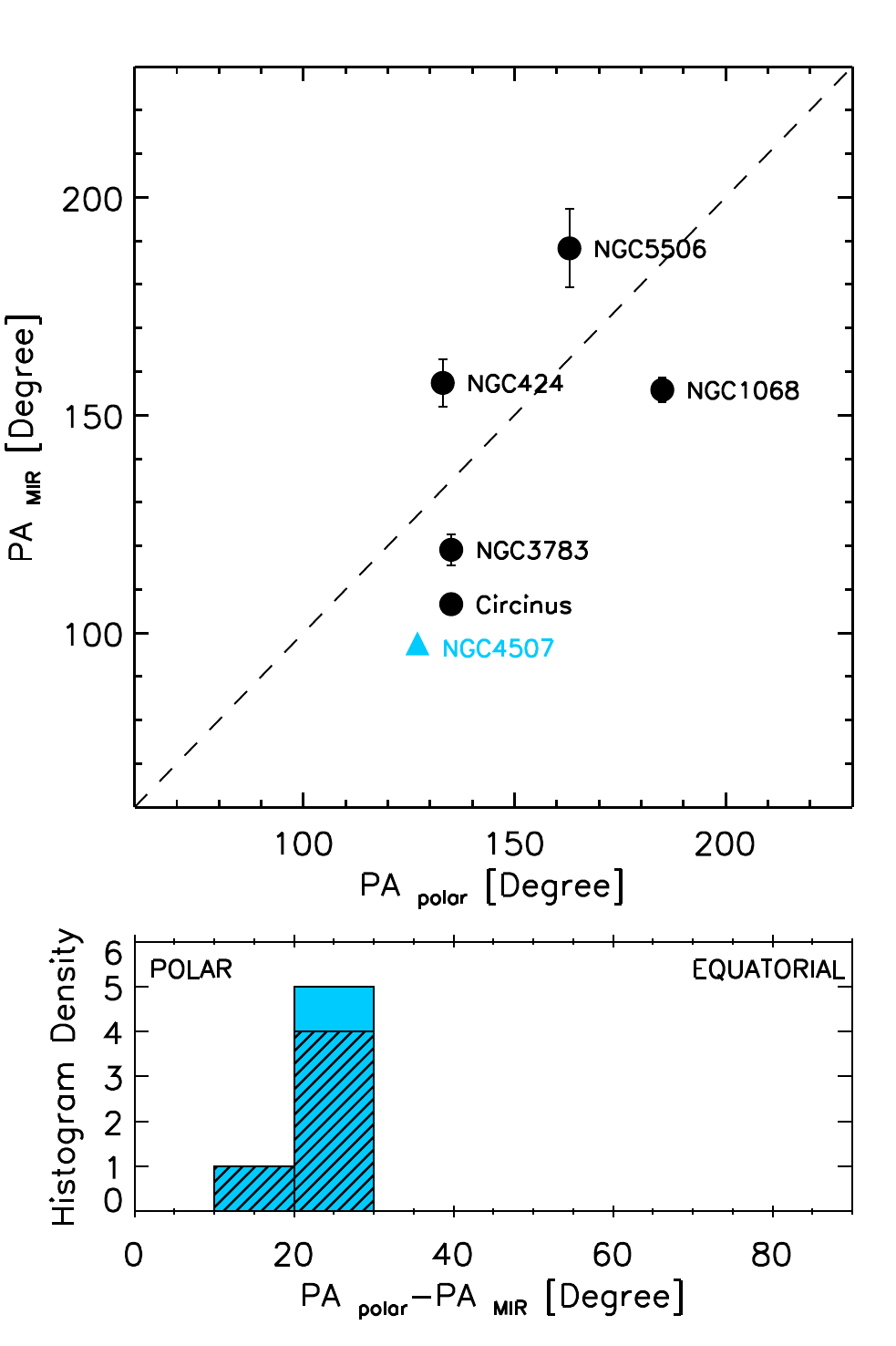}
   \captionof{figure}{{\it Top)} Comparison between the position angle of the mid-infrared emission from the parsec-scale structure ($PA_{MIR}$) and the inferred position angle of the system polar axis ($PA_{polar}$) for the elongated objects. 
   For completeness we also include the marginally elongated object NGC4507 (blue triangle), no errorbars are determined for this object.
   The dashed line represents a one to one relation for the position angles.
   {\it Bottom)} Histogram of the difference between the inferred system polar axis and the mid-infrared position angle obtained using interferometry.
   The histogram including the marginally elongated object NGC4507 is shown in blue bars, while  elongated objects are given with black bars. 
   Only objects with elongations obtained from interferometric data have been used here.}
   \label{fig:polar}
\end{figure}

We observe that in all five objects, where we report an elongation, the position angle of the major axis of the  mid-infrared emission is always closer to the inferred polar axis of the system than perpendicular to it, see Fig~\ref{fig:polar}. 
If we assume a marginally elongation for the object NGC 4507 to be true, then the mid-infrared PA for this object is also close to the polar system axis.
Additionally, we observe that if the estimates of the polar axis system are indeed representative of the true polar axis system, then there is a lack of axial symmetry of the mid-infrared emission around the polar axis (see bottom plot of Fig~\ref{fig:polar}).
In all the objects the emission is always leaned slightly towards one of the sides of the polar axis. 
In the Circinus galaxy, for example, the difference is 20-30 degrees and in this source the positions angles of both the mid-IR emission and the ionization cone are very reliably determined.

Additionally, the multiple infrared shapes (Elongated: 1 Type I, 3 Type 2s, and 1 Type 1i and nearly circular: 1 Type 2 and 1 Type 1i) almost regardless of the Seyfert Type may be a surprising result if one has a picture in mind of a ``donut''-like torus \citep[e.g. Fig. 1 of][]{1993ARA&A..31..473A}.

It has been long realized, however, that such a simple structure cannot represent the actual distribution of dust.
Instead, the dust must be in a clumpy configuration \citep[e.g.][]{2008A&A...482...67S} whose image, as simulated using radiative transfer, is complex and its reduction to an ``elongated Gaussian model'' is not at all obvious. 
If also taking into account hydrodynamical effects, even a more or less azimuthal dust configuration can produce images that are polar-elongated 
(e.g. Fig. 8 of \citealt{2009MNRAS.393..759S} and Fig. 5 of \citealt{2014MNRAS.445.3878S}).
In fact, models that produce an infrared emission with a X-shaped morphologies \citep[e.g.][]{2005A&A...437..861S} also show polar extension to the zero-th order. 
Slightly asymmetries in the density distribution produced by filaments or clouds could then explain the lack of asymmetry around the polar axis system.
Additional information, such as kinematics, would be needed to distinguish such a scenario from, e.g. a disk wind scenario which would also produce polar-elongated emission  \citep[e.g.][]{2012ApJ...758...66W, 2013arXiv1303.7142G, 2014MNRAS.445.3878S}.

While with infrared ``images'' alone we cannot provide a complete panorama about the structure of the torus, our result of the polar extended emission should serve as a constraint for dusty models that attempt to provide a description of the dusty environment in AGNs.
Current SED fitting studies do not take this into account as the SED does not provide any geometrical information.
Further investigation needs to be done to see if derived torus properties, such as covering factors, torus sizes, cloud numbers from torus models that reproduce the polar-like extended emission are consistent with the current models used by the community.

\subsection{Future improvements}

In our reliability analysis we studied how the ($u,v$) coverage and the signal-to-noise ratio influence our ability to measure an intrinsic elongation.
From this study we learned that the most relevant to determine an elongation is the filling of the ($u,v$) coverage.
Without improving the accuracy of the measurements but just by obtaining more equally distributed ($u,v$) points at different position angles, we would be able to confirm or rule out elongations in at least 6 objects (LEDA17155, NGC~4151, NGC~4593, IC4239A, NGC~5995, and NGC~7469).
These objects have relatively low to intermediate point source fractions (< 0.6) but their current ($u,v$) coverage typically has only measurements, at resolutions where the emission is resolved, along two distinct position angles.  
For the objects IZwicky1, H0557-385, NGC~3281, due to their relatively high point source fractions $\sim 0.8$ an improvement of both the ($u,v$) coverage and the signal-to-noise ratio  would be needed to learn more about the mid-infrared geometry of these objects.
It is worth noting that for the objects NGC~4151, NGC~4593 and NGC~3281 there is evidence for a 100 pc scale elongated structure with orientation also close to the polar axis system \citep{2014MNRAS.439.1648A}. 
For these three objects, new infrared interferometric observations would be required to filled the missing ($u,v$) points and be able to investigate the nature of the elongated emission.

The remaining objects NGC~1365, Mrk~1239, 3C273, ESO323-77 and NGC5128 (Cen A) and other objects with point source fractions $\approx 1$ require longer baselines to resolve their compact mid-IR emission.
Results from our reliability analysis should be used to plan future observations with the second generation instrument MATISSE \citep{2014Msngr.157....5L} to improve our view of the mid-infrared emission in AGNs.

\section{Conclusions}

We have analyzed the mid infrared interferometric data of 23 sources observed with the instrument MIDI at the VLTI and investigated in how many objects we should be able to find elongations given the observed $(u,v)$ coverages along with the achieved signal/noise ratios and the minimum visibilities.
We found that we should be able to find elongated emission in seven out of 23 objects and indeed in five of these seven elongated emission is found (four objects are already published, NGC~5506 is a new detection). 
However, we also find two sources (NGC4507 and MCG-5-23-16) which are compatible with non-elongated, i.e. circular emission.
In the sources where deviations from a circular emission geometry are detected, the elongation is always much closer to the polar than the equatorial direction of the system.  The trend of polar elongated emission at parsec scales is compatible with what is observed at 100 pc scales (e.g, NGC4593, NGC3281, \citealt{2014MNRAS.439.1648A}; NGC4151, \citealt{2003ApJ...587..117R}; Circinus, \citealt{2003ApJ...587..117R}; NGC 1068, \citealt{2005MNRAS.363L...1G}).

This needs to be followed up with higher signal/noise and more homogeneous $(u,v)$ coverage before further conclusions can be drawn.
Most importantly, a complete sample of AGN tori \citep[e.g. selected by hard X-rays such as][]{davies2015} should be observed with the upcoming four-beam mid-IR interferometer MATISSE at the VLTI \citep{2014Msngr.157....5L} in order to assess the shape and sizes of AGN tori in a representative sample.

\begin{acknowledgements}
The authors thank the anonymous referee for the thoughtful and helpful comments. 
N. L\'opez-Gonzaga was supported by grant 614.000 from the Nederlandse Organisatie voor Wetenschappelijk Onderzoek and acknowledges support from a CONACyT graduate fellowship.
L. Burtscher is supported by a DFG grant within the SPP 1573 ``Physics of the interstellar medium''.
\end{acknowledgements}

\bibliographystyle{aa} 
\bibliography{/home/nlopez/Documents/PhD_thesis/In_use/biblio/reference_thesis}

\end{document}